\documentclass[sigconf]{acmart}
\AtBeginDocument{%
  }

\copyrightyear{2026}
\acmYear{2026}
\setcopyright{cc}
\setcctype{by}
\acmConference[CHI '26]{Proceedings of the 2026 CHI Conference on Human Factors in Computing Systems}{April 13--17, 2026}{Barcelona, Spain}
\acmBooktitle{Proceedings of the 2026 CHI Conference on Human Factors in Computing Systems (CHI '26), April 13--17, 2026, Barcelona, Spain}
\acmPrice{}
\acmDOI{10.1145/3772318.3791834}
\acmISBN{979-8-4007-2278-3/2026/04}

\usepackage{stfloats}
\usepackage{colortbl}
\newcommand{\datasetname}{\textsc{Ask}\&\textsc{Prompt}}
\newcommand\revised[1]{\textcolor{black}{#1}}

\begin{document}

\title{Say It My Way: Exploring Control in Conversational Visual Question Answering with Blind Users}


\author{Farnaz Zamiri Zeraati}
\authornote{Both authors contributed equally to this research.}
\orcid{0000-0003-2986-0664}
\affiliation{%
  \institution{University of Maryland}
  \city{College Park}
  \state{Maryland}
  \country{USA}
}
\email{farnaz@umd.edu}

\author{Yang Trista Cao}
\authornotemark[1]
\orcid{0009-0000-5296-3229}
\affiliation{%
  \institution{University of Texas at Austin}
  \city{Austin}
  \state{Texas}
  \country{USA}}
\email{ycao55@utexas.edu}

\author{Yuehan Qiao}
\orcid{0009-0007-9410-7317}
\affiliation{%
  \institution{University of Maryland}
  \city{College Park}
  \state{Maryland}
  \country{USA}
}
\email{yhqiao@umd.edu}

\author{Hal Daumé III}
\orcid{0000-0002-3760-345X}
\affiliation{%
  \institution{University of Maryland}
  \city{College Park}
  \state{Maryland}
  \country{USA}
}
\email{hal3@umd.edu}

\author{Hernisa Kacorri}
\orcid{0000-0002-7798-308X}
\affiliation{%
  \institution{University of Maryland}
  \city{College Park}
  \state{Maryland}
  \country{USA}
}
\email{hernisa@umd.edu}







\begin{abstract}
\revised{Prompting and steering techniques are well established in general-purpose generative AI, yet assistive visual question answering (VQA) tools for blind users still follow rigid interaction patterns with limited opportunities for customization. User control can be helpful when system responses are misaligned with their goals and contexts, a gap that becomes especially consequential for blind users that may rely on these systems for access.} We invite 11 blind users to customize their interactions with a \revised{real-world} conversational VQA system. Drawing on 418 interactions, reflections, and post-study interviews, we analyze prompting-based techniques participants adopted, including those introduced in the study and those developed independently in real-world settings. VQA interactions were often lengthy: participants averaged 3 turns, sometimes up to 21, with input text typically tenfold shorter than the responses they heard. Built on state-of-the-art LLMs, \revised{the system} lacked verbosity controls, \revised{was limited in estimating distance in space and time,} relied on inaccessible image framing, and offered \revised{little to} no camera guidance. We discuss how customization techniques such as prompt engineering can help participants work around these limitations. Alongside a new publicly available dataset, we offer insights for interaction design at both query and system levels.

\end{abstract}





\begin{teaserfigure}
    \centering
    \includegraphics[width=\textwidth]{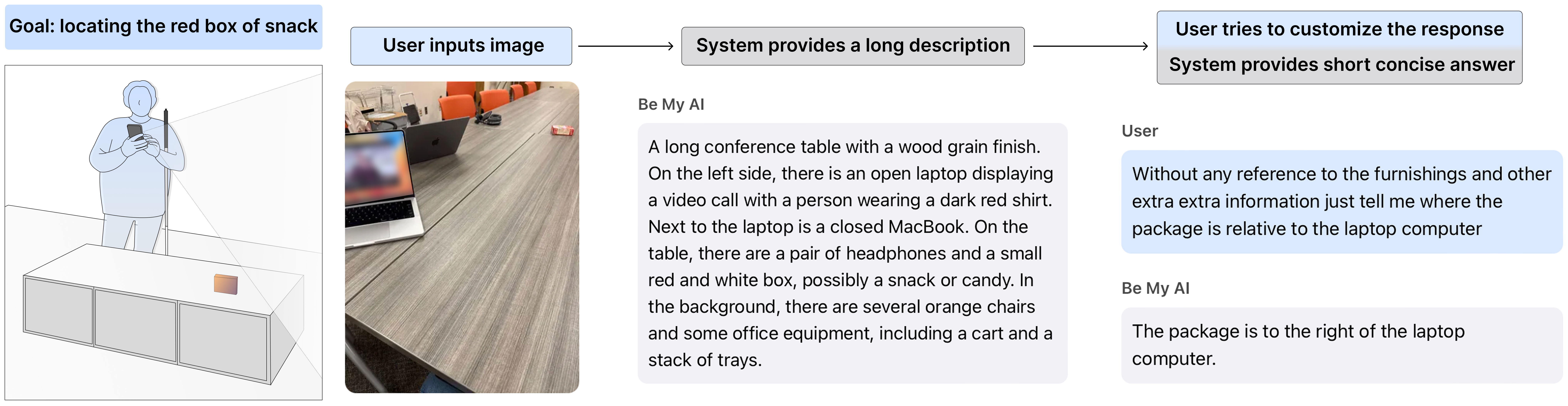}
    \caption{During an in-lab scenario, P11 experimented with stylistic customization to control the level of detail in the system’s description and focus the response on locating the snack box.}
    \label{fig:teaser}
    \Description[]{A blind participant capturing a lab room trying to find the red box of snack, first the application provides a long description of what is included in the scene, after that the participant tries to filter out extra information by asking the app to only provide the red box location relative to laptop, as a result the answer is short and concise and focused on the goal.}
\end{teaserfigure}

\maketitle

\section{Introduction}
\revised{Visual question answering (VQA) aims to answer open-ended questions about images using natural language responses and has long been explored as a way to help blind people access  visual surroundings and content. Early approaches relied on sighted assistance, such as crowd-sourced and real-time services, who could help localize objects and describe scenes~\cite{bigham2010vizwiz, lasecki2013Answering, bemyeyes, aira, bigham2010locateit, burton2012crowdsourcing}. Advances in computer vision and natural language processing have shifted many of these approaches to automated systems that generate descriptions and answer visual questions, promising greater privacy and independence~\cite{s2024visqaid, LE2021multi, Fernandes2024VIIDAAI, Singh2019TowardsVM, Mohith2020Visual}.
More recent developments in generative AI and vision–language models (VLMs) have enabled conversational, multi-turn interpretations of visual scenes, powering real-world applications such as the \textit{Be My AI} feature in \textit{Be My Eyes}~\cite{ bemyeyes, Oliver2024how}.}

\revised{Yet relatively little attention has been given to how blind users can shape or control these increasingly capable systems. Existing applications typically} follow a fixed interaction pattern: a user provides an image, receives a lengthy system-generated description, and may then ask follow-up questions. While functional, this design limits the degree of control blind users can exercise and the extent to which responses align with their goals, preferences, and contexts. This gap is especially important because blind users may rely on these systems in situations where efficiency and clarity are critical~\cite{Gonzalez2024Investigating}. A lengthy or poorly targeted response can create confusion, delay action, or even pose safety risks~\cite{Hong2024Understanding}. \revised{This is in contrast to general-purpose generative AI applications, where prior work documents a wide range of approaches, such as prompt engineering~\cite{Zhijie2024PromptCharm, Stephen2023Promptify, Yingchaojie2023PromptMagician} and fine-tuning models with personal data~\cite{Zheng2019PersonalizedDG,userlibri_2022}, that let  users influence the type, tone, and depth of information they receive.}

\revised{Motivated by this gap, we investigate the following research question: \textit{What are the opportunities and challenges for blind users to exert control over VQA responses?} }
\revised{We} conduct a three-phase study examining how blind users interact with existing customization techniques for controlling AI-generated responses in conversational VQA systems. The first phase is a controlled in-lab session, where participants (N=11) are introduced to and practice a set of customization techniques, such as binary feedback, zero-shot style prompting, and chain-of-thought prompting, through predefined tasks (see \autoref{fig:teaser} for an example). The second is a  diary study. For 10 days, participants use the system independently in their daily life, documenting naturally occurring interactions across diverse tasks and environments. In the last phase, participants join a remote semistructured interview, where they share deeper insights on their interactions, preferences, and reflections on system behavior and control. This combined approach of short in person sessions with extended multi day engagement aims to capture both immediate, guided user behaviors and longitudinal, naturally occurring interactions where customization techniques may emerge, get adapted, and evolve over time. With the reflections we aim to understand how VQA systems can better support user control beyond the prompt engineering paradigm.

Our study reveals that interactions are often lengthy and imbalanced: participant inputs are typically concise, while system responses are more than ten times longer. Conversations commonly stretch over several turns (median = $3$, up to $21$), especially when participants make multiple attempts to capture usable images, clarify ambiguous outputs, or pursue curiosity-driven questions. While most interactions are successful, failures arise from inaccessible or missing information in the image, system limitations, or model errors.
Participants actively experiment with ways to shape responses. Many adopt customization techniques introduced in the lab---such as binary feedback and targeted prompting---while also developing new strategies in the diary study. Customization is most common in professional contexts but also appears in familiar and unfamiliar environments, with slightly higher rates in familiar settings. Through these practices, participants seek to manage verbosity, direct attention to task-relevant details, and strengthen trust in system output. Reflections across tasks highlight further needs: toggling verbosity based on time pressure, spatial descriptions grounded in user-centered reference points rather than image frames, proactive support for image capture, and verification mechanisms for high-stakes situations.

This work makes two contributions: (1) empirical findings showing how blind users adapt to current conversational VQA constraints while highlighting design opportunities for greater control to support efficiency, clarity, and trust, and (2) the \datasetname~dataset, which contains images, text conversations from our study with the \textit{Be My AI} application, and contextual annotations \revised{from our research team grounded on participant feedback}.





\section{Related Work}
\subsection{Personalization in Generative AI}
% ...

Personalization of large language models (LLMs) has been extensively studied across a range of natural language processing tasks, including dialogue systems \cite{dialogue_related_1,dialogue_related_2}, summarization \cite{summarization_related}, and search and information retrieval \cite{search_related_1,search_related_2}. As commercial and publicly accessible LLMs become more widespread, personalization is increasingly viewed as a key solution to several emerging challenges--enabling more usable systems across diverse tasks \cite{yun_enhancing_2024,ouyang_training_2022}, improving alignment with individual user values \cite{gabriel_AI_2020,blair2024democratizingrewarddesignpersonal}, and fostering greater user autonomy~\cite{Kirk_Vidgen_Röttger_Hale_2024,Lacroix2023WhosIC}. In response, researchers have proposed a range of approaches to enable personalization at various stages of the machine learning pipeline. During pretraining and fine-tuning, datasets have been proposed to better represent diverse user groups, tasks, and values to ensure broader inclusivity in foundational model behavior~\cite{Zheng2019PersonalizedDG,userlibri_2022,liu_JDsearch_2023}. Techniques such as supervised fine-tuning (SFT) and reinforcement learning from human feedback (RLHF) \cite{ouyang_training_2022} have also been developed to steer models toward more desirable, value-aligned outputs~\cite{Eleanor2025Choice, Jin2024Orchestrating, Minhyeon2024Active, Jianfei2024Disentangling, Yuanpu2024Personalized}.
In the post-training phase, studies have investigated the effectiveness of few-shot prompting and in-context learning to adapt model behavior without requiring parameter updates--allowing users to shape outputs using lightweight, example-based techniques~\cite{zhao2024grouppreferenceoptimizationfewshot,li-etal-2024-steerability,salemi-etal-2024-lamp,hwang-etal-2023-aligning}. Building on this, recent work investigates user-facing customization at inference time, including interactive tools that support prompt exploration and output refinement through mixed-initiative designs, visual aids, and LLM-powered suggestions ~\cite{Zhijie2024PromptCharm, Stephen2023Promptify, Yingchaojie2023PromptMagician}. These systems have been shown to reduce cognitive load, accelerate creative workflows, and improve alignment between prompts and system responses \cite{Stephen2023Promptify, yuhan2024PrompTHis}. Additionally, newer interfaces move beyond text, incorporating input modalities such as annotations, scribbles, and visual elements to enhance accessibility and user control \cite{John2023PromptPaint, Hyerim2025Exploring, Seungho2023PromptCrafter}.
Extending beyond text, personalization also was explored with visual language models (VLMs) and multimodal systems \cite{userlibri_2022,alaluf_myvlm_2024,nguyen2024yollava}.
In this study, we focus on the VLM-based visual question answering system to explore how blind users take control and customize the systems to their needs and preferences.

\subsection{Blind Users and Generative AI}
There is a rich literature on leveraging AI to benefit the blind community. Surveys highlight applications across text access, perception and description, and mobility, delivered through smartphones, wearables, and head-mounted cameras e.g,~\cite{Bhanuka2023what, Shrinivas2023impact, Khansa2024Digital}. While these efforts show steady progress, reviews point out that research has focused heavily on description tasks~\cite{Bhanuka2023what, Mohammad2022recent}, and that concerns around evaluation, privacy, and safety remain~\cite{zahra2024reporting, julia2025technology}.

Researchers and developers have been quick in applying generative AI to assist blind users. VLMs such as \textit{Be My AI}~\cite{bemyeyes} among others turn photos into conversational descriptions and are being integrated into mobile apps~\cite{Francisco2024development} and wearables~\cite{Bal2024Exploring, Mirza2024AIbasedWearable, batyr2023image} for real-time access to visual scenes, objects, and text~\cite{Hafeth2023Cloud, Alqahtani2023Ayn}. For navigation, prototypes combine LLM reasoning with on-device sensing to offer step-by-step guidance and hazard detection~\cite{zhang2024enhancing, Bal2024Exploring, Kim2025Toward, Sangmim2024guide}. Evaluations highlight persistent challenges such as hallucinations, limited contextual awareness, and difficulty supporting goal-directed interaction~\cite{heidi2024assisting, Evaluating2025Antonia}. User studies show both enthusiasm and hesitation: blind users experiment with these tools in household, social, and safety contexts, but often develop their own verification strategies to manage uncertainty~\cite{Adnin2024Ilook, xie2024emergingp, Gonzalez2024Investigating, ricardo2025towards}. Together, this literature points to opportunities for more reliable and personalized multimodal solutions.

Within this growing space, personalization for assistive generative AI has been explored through two complementary directions: interaction and interface design, and model-level adaptation. Examples of personalization through interaction and interface design are systems that adjust descriptions based on user intent, scene context, or verbosity~\cite{Ruei2024worldscribe, ananya2024context}. Closer to model-level adaptation, projects on teachable object recognition have shown how users can train systems to handle personally relevant items~\cite{findmything, kacorri2017people, Sosa2017hands}. Our study builds on the first direction by examining prompting and customization techniques across both controlled lab tasks and real-world diary studies, exploring a broader view of how blind users tailor VLM-based conversational VQA system to their needs.

\section{Methods}
\begin{figure*}[t]
    \centering
    \includegraphics[width=\textwidth]{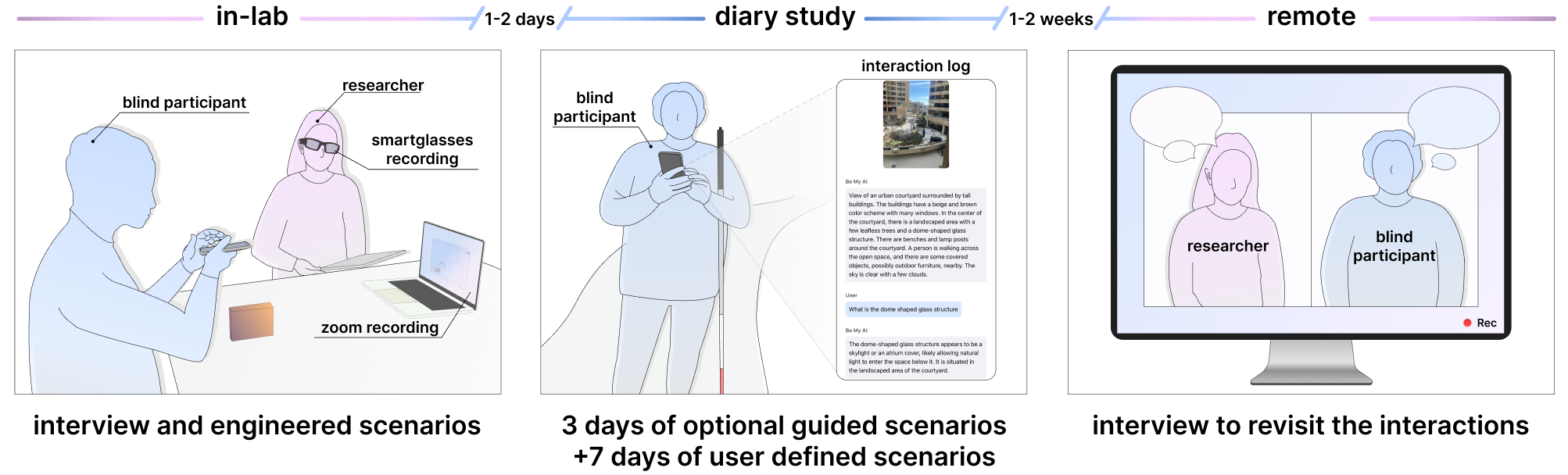}
    \caption{Our multi-day study includes an in-lab interview with scenarios, a 10-day diary, and a remote interview.}
    \label{fig:study}
    \Description[]{artistic depiction of the study, from left to right: in-lab interview and engineered scenarios, the diary study with user-defined scenarios, and the remote interview. The first illustrates a blind participant taking a photo of an object on top of a table using the phone. The participant is seated close by a researcher, who is wearing a pair of smart glasses for video recording.  An open laptop placed in front of the participant on the table is also recording via Zoom. The second illustrates a blind participant with a white cane using a smartphone which displays an image of an urban courtyard and a text description followed by a user question and AI answer about a dome-shaped glass structure. Under this section there is text specifying that this phase included 3 days of optional guided scenarios plus 7 days of user defined scenarios. The third illustrates A computer screen showing a Zoom video call with the researcher and blind participant speaking, indicated by speech bubbles and a red “Rec” icon.}
\end{figure*}

To explore how blind users interact with AI-powered conversational VQA systems and attempt to control system responses, we conduct a multi-day user study (\autoref{fig:study}). Participants first visit our lab for a semi-structured interview, an introduction to the commercial VQA app \textit{Be My AI}, and 5 carefully engineered interaction scenarios. They then complete a diary study, using \textit{Be My AI} in daily life and sharing at least 25 interaction scenarios with their goals and reflections. Last, participants join a remote semi-structured interview to reflect on their diary entries.

\subsection{Recruiting Participants}

We recruited participants through three channels: our lab's mailing list, the National Federation of the Blind (NFB)~\cite{nfb2026}, and word of mouth. The mailing list included blind individuals who had previously participated in our research and consented to future contact. Those in the list and the NFB received an email with a study description and consent form. Interested individuals contacted us by email and were screened for eligibility. To participate, individuals had to be 18 or older and identify as blind. All participants provided informed consent and were reminded that participation is voluntary and can be withdrawn at any time, with compensation provided for completed portions. A total of $13$ participants enrolled; two withdrew after the in-lab session (one citing time conflicts), leaving $11$ for analysis. Of those, four identified as totally blind and seven as legally blind (Table \ref{tab:participant}). The average age was $42$ years (SD $= 11.33$). Nine participants identified as women and two as men. \revised{Education levels ranged from high school to a JD degree; employed participants worked in management, technology, customer service, or law. Most participants reported being somewhat familiar with AI, with a broad understanding of what it does, and none reported extensive knowledge. Most participants had used \textit{Be My Eyes} (except P5), though fewer than half mentioned \textit{Be My AI}. Many also used \textit{Seeing AI} and \textit{Aira}. A few had tried \textit{Meta Glasses}, and one mentioned \textit{ChatGPT}.} Participants received a fixed compensation of $\$300$, based on an estimated 15 hours of study activities at a \$20/hour, including 2.5 hours for the in-lab session, 25 minutes of sharing and 35 minutes of reflection per diary day, and 2.5 hours for the final interview. This rate reflected expected effort rather than incentivizing excessive engagement, and no participant exceeded the allotted time for the in-lab or remote interviews we could track. 
The study protocol was approved by our institution’s review board (IRB \#2160619-2).

\begin{table*}
\small
\caption{Self-reported participant demographics, familiarity with AI, \revised{and related assistive technology use}.}
\small\renewcommand{\tabcolsep}{2pt}
\resizebox{\linewidth}{!}{
\begin{tabular}{cccccccc}
\toprule
\textbf{PID}& \textbf{Blindness} & \textbf{Age} & \textbf{Gender} & \textbf{Education}& \textbf{Occupation} & \textbf{AI Fam*} & \revised{\textbf{Related Assistive Tech}} \\
\midrule

P1 & total & 30 & W & associate & full-time parent & 1 & \revised{Be My Eyes} \\
\rowcolor{gray!20}
P2 & legal & 39 & W & high school & customer service representative & 3 & \revised{Seeing AI; Be My Eyes; Aira} \\
P3 & total & 52 & W & some college & not employed & 3 & \revised{Be My Eyes; Seeing AI} \\
\rowcolor{gray!20}
P4 & total & 38 & W & high school & not employed & 2 & \revised{Be My Eyes} \\ 

P5 & legal & 53 & W & master & not employed & 3 & \revised{-} \\   
\rowcolor{gray!20}
P6 & legal & 31 & W & some college & assistant technology tutor & 3 & \revised{Be My AI; Aira; Meta Glasses} \\  

P7 & legal & 34 & M & master & budget manager & 3 & \revised{Be My AI; Be My Eyes; Seeing AI} \\   
\rowcolor{gray!20}
P8 & legal & 42 & W & master & management specialist & 2 & \revised{Be My Eyes; Seeing AI; Alexa} \\  

P9 & legal & 25 & W & bachelor & accessibility program manager & 3 & \revised{Be My Eyes; Aira; ChatGPT; Seeing AI}\\     
\rowcolor{gray!20}
P10 & legal & 56 & W & master & manager of a care program & 3 & \revised{Seeing AI; Be My AI; Aira} \\  

P11 & total & 57 & M & juris doctor & lawyer & 3 & \revised{Aira; Be My AI; Meta Glasses} \\     
\bottomrule
\end{tabular}
}
\begin{minipage}{\textwidth}
    \smallskip
    \revised{*Familiarity with AI was reported in a} 4-point scale: 1 = not familiar (never heard of it), 2 = slightly familiar (heard of it but don’t know what it does), 3 = somewhat familiar (broad understanding of what it is/does), 4 = extremely familiar (extensive knowledge).
\end{minipage}
\label{tab:participant}

\end{table*}


\subsection{Introducing Customization in the Lab}

\begin{table*}
\small
\caption{Definitions and examples of the customization techniques introduced in the lab.}
\renewcommand{\arraystretch}{1.4} 
\centering
\begin{tabular}{p{7em} p{13em} p{16em} p{11em}}
\toprule
\textbf{Technique} & \textbf{Definition in GenAI}
& \textbf{Definition in our use case} & \textbf{Example used} \\
\midrule
Binary Feedback & Feedback that indicate positive or negative rewards \cite{ouyang_training_2022}. & Users provide a simple positive or negative feedback with the response & \textit{"Helpful/Not helpful."} \\
\midrule
Zero-shot: Style Prompting & Prompt to specify desired stylistic attributes of GenAI output (e.g., tone and genre) \cite{lu2023boundingcapabilitieslargelanguage}. & Users prompt to specify desired style to adjust following responses. & \textit{"More/Less detail."} \\
\midrule
Zero-shot: Intention Prompting\textsuperscript{*} & Emotion prompt that incorporates phrases of psychological relevance to humans \cite{li2023largelanguagemodelsunderstand}. & Users express their goals to guide systems tailor responses based on the task context.  & \textit{"I want to gain muscle. Is this a good option of food?"} \\
\midrule
CoT Prompting & Offer reasoning behind to guide GenAI systems \cite{patel-etal-2022-question,cot}. & Users prompt with reasoning for desirable or undesirable answers.  & \textit{"I only care about finding the soup box, so do not mention about other items."}\\
\midrule
Image-as-text Prompting & Generate a textual description of image(s), enabling easy inclusion into a text-based prompt\cite{hakimov2023images}. & Users prompt with visual characteristics of objects or scenes, allowing systems to complete tasks with additional information.  & \textit{"Find my keys with a keychain that is green, round, and with raised letters."} \\
\bottomrule
\end{tabular}
\begin{minipage}{0.96\textwidth}
    \smallskip
    \textsuperscript{*}Zero-shot: Intention prompting is generalized from emotion prompting.
\end{minipage}

\label{tab:customization_techniques}
\end{table*}
 
During the lab visit, participants were introduced to the conversational VQA system used throughout the study: \textit{Be My AI}, a feature of the \textit{BeMyEyes} mobile application\footnote{We designed the study using \textit{Be My AI} v6.0. During the study, participants used v6.0-6.3, which differed only in minor bug fixes and language support.}. This phase aimed to (1) ensure participants were comfortable navigating the app and (2) expose them to existing customization techniques while gathering initial perspectives. We selected \textit{Be My AI} because, at the time, it was the only mature, widely available, and screen-reader accessible AI-powered visual assistance tool. In-lab sessions were recorded with two devices: a laptop camera capturing the front view and a pair of smart glasses capturing the researcher's  view of the environment and participant interactions.

During the session, we first collected participants’ demographics and their attitudes toward and experience with technology. This was followed by a guided practice session where participants engaged with the system across five scenarios simulating  real-world tasks. These scenarios were informed by frequent use cases from the VizWiz dataset \cite{vizwiz}, prior work with blind users \citep[e.g.,][]{findmything,Islam_2024,Tang_2025,bigham2010locateit}, and discussions with a blind researcher.

We introduced a set of customization techniques, \revised{appropriate for these scenarios}, grounded in established  prompting \revised{techniques} for large language models. Drawing on the taxonomy by ~\citet{schulhoff2025promptreportsystematicsurvey}, we selected strategies relevant to visual assistance and adapted them to fit short, mobile-based interactions and existing functionalities in \textit{Be My AI}. A full list of introduced techniques appears in \autoref{tab:customization_techniques}, with detailed scenarios in \autoref{appendix:A}. To our knowledge no prior work has systematically documented prompting \revised{techniques} for accessibility. We present our adaptation as an initial step toward such a framework. However, because \textit{Be My AI} does not support persistent customization, participants did not experience lasting effects. To manage expectations, we informed them of this limitation but encouraged them to use the techniques as if preferences could carry over.
After each scenario, participants reflected on goal achievement, ideal responses, and customization techniques, comparing them and noting preferences across contexts.

\subsection{Collecting Diary Data in Everyday Contexts}
A day or two after the in-lab session, participants began a 10-day diary study using \textit{Be My AI} independently in their everyday environments. The first 3 days were lightly structured to help participants become comfortable with the system and to encourage early exploration across contexts. From Day 4 through Day 10 there was no prompting beyond submitting at least 25 interactions, experimenting with different ways of guiding the system including the customization techniques introduced in the lab. Instructions were sent via email or message, according to preference. For each interaction, participants shared a brief description and link, noting the task goal, environment familiarity, and time constraints. To support deeper reflection, they also completed a short set of end-of-day questions. \revised{We note that while our data collection focused solely on interactions submitted through \textit{Be My AI}, participants were not restricted in their technology use and were free to use any other assistive tools throughout the diary study.} Full protocol details for Days 1-3 and reflection questions appear in \autoref{appendix:B} and \autoref{appendix:C}.

\subsection{Conducting Post-Diary Interviews}

Within two weeks of the diary study, participants completed a virtual interview on their preferred platform (Zoom or Microsoft Teams). The goal was to gain deeper insights into their interactions, preferences, and reflections on system behavior and control. With consent, interviews were audio-recorded and transcribed.

The interview began with scenario confirmations. We selected submitted diary interactions that lacked context---such as goals, time sensitivity, familiarity with the setting or object, or task outcome---and asked participants to confirm or correct our interpretation. We also asked whether any interaction raised privacy concerns and obtained explicit consent for those they agreed to share for future research and publications.

Next, we reviewed five to six  interactions per participant, selected to represent diverse tasks. For each, participants described their ideal system response in terms of content (e.g., type of information) and style (e.g., tone, level of detail), and whether their preferences would differ across contexts, such as when they were in a hurry or not.

We then explored customization techniques. Participants reflected on why they used a technique and how effective it felt, or if they had not, why not and what alternatives they could imagine. Discussions revealed not only how they customized the system, but also which techniques felt intuitive, accessible, or worth the effort in other contexts.

Finally, we gathered reflections on overall experience. Participants rated two usability statements---ease of use and comfort---on a 5-point Likert scale and answered open-ended questions about customization techniques, including which they preferred and which felt challenging or unnatural. They also reflected on future use: when they might trust the system without customization and when they would invest effort to guide or verify responses.

\subsection{Analyzing Data Across Sessions}

To better understand how blind participants interacted with the system and adopted customization techniques, we triangulated data from multiple sources: recordings from the in-lab and post-diary interviews, interaction logs from the diary study, and contextual notes participants provided alongside each interaction.

\subsubsection{Participant feedback during in-lab session}
 We employed inductive thematic analysis~\cite{huberman2014qualitative}, where transcriptions were first organized by the questions.  The research team then added interpretative codes and notes through iterative discussions: first establishing a coding framework on a 20\% sample, then applying and refining it across the remaining data. This process produced overarching themes capturing participants’ expectations of ideal responses, preferences for customization techniques, and strategies for validating responses.


\subsubsection{Annotation of diary study interactions}
Participants shared links to their interactions, which included images and chats, and provided context. We annotated these data using deductive thematic analysis \cite{doi:10.1177/160940690600500107} and content analysis \cite{harwood2003overview}. Predefined aspects included setting (indoor vs. outdoor; professional vs. leisure), familiarity with the environment/object/people, time sensitivity, and whether participants were alone or with others. \revised{We divided the} annotation among three researchers. \revised{Before the post-interview, researchers pre-populated only those annotations that could be unambiguously inferred from the interaction itself (image and text) or from participant-provided context (e.g., clear indoor/outdoor cues; explicit references such as “my house” or “my ring doorbell” for annotating environment or item familiarity). Annotations that required information about participants’ internal states or social context, such as whether they were in a hurry or whether others were present at the time of interaction, were intentionally left blank and were completed during the post-interviews by asking participants directly.}
We also coded the types of participant questions to the system, finding strong parallels with the categorization scheme of \citet{Brady2013investigating}, where questions were instead answered by humans not AI. 
Last, we coded the customization techniques used. While most aligned with  categories introduced in the lab,  new techniques also emerged, which we  categorized following the taxonomy in \cite{schulhoff2025promptreportsystematicsurvey} \revised{(see Figure~\ref{fig:taxonomy})}.

\subsubsection{Participant feedback during post-diary interview}
 \revised{Before the post-interview, one researcher reviewed all participant-submitted interactions for any potentially personally identifiable information (PII).} The researcher also selected 5–6 diverse and discussion-provoking samples, which were later grouped by question type.
 
 \revised{During the interview, we walked through all interactions one by one. For each interaction, the researcher informed the participant of any identified PII and explicitly asked whether they consented to share the interaction—including images, user prompts, and system responses—either in full, in redacted form, or not at all.} We also refined and updated the diary annotations, \revised{adding contextual details that could not be inferred beforehand (e.g., whether the participant was in a hurry, whether others were present). Finally, the participant reflected on the selected interaction samples in terms of ideal responses, customization techniques, and comparing them across contexts and noting how their preferences shifted depending on the task or environment.}
 
 After the interview, we grouped the transcribed samples by participant and applied inductive thematic analysis using the same approach as in the in-lab study.

\section{Results}

From the in-lab session, we collected a total of $55$ \textit{Be My AI}  interactions (five per participant) and $1,041$ minutes of researcher-perspective video recorded via smartglasses (58:36 -- 125:42, SD = 22:51), backed up by Zoom recordings via a laptop. 
In the diary study, we collected $363$ interactions ($22$–$48$ per participant), typically accompanied by brief text message descriptions context and user goal. Participants often interwove daily reflections with these entries. We also recorded $1,232$ minutes of post-study interviews  in which participants discussed their experience in depth. In the \datasetname~dataset available at \url{https://iamlabumd.github.io/ask-prompt/}, we share $408$ \revised{(418-10)} interactions from the lab and diary study with our annotations. Ten interactions were omitted at participant's request, and $43$ images/texts were redacted to remove personally identifiable information. Each interaction is manually annotated with details such as customization techniques, context, and user goal.

\subsection{Characteristics of the Interactions}\label{sec:4.1}

Efficiency is a critical factor in assistive systems, particularly for blind users who already face time disparities when navigating accessibility barriers in both physical~\cite{zhang2019effect} and digital~\cite{griffith_quantifying_2023} contexts. AI-powered VQA systems promise faster, near-instantaneous responses compared to volunteer sighted assistance, for more efficient, real-time access to visual information. Customization further enhances this potential by allowing users to tailor interactions to achieve their goals quickly~\cite{Kirk_Vidgen_Röttger_Hale_2024}. 
However, estimating the time blind users spent per interaction in our study is challenging. Input methods vary (typing vs. dictation), each with different speeds \cite{hong_speech_2020, Alnfiai2024ExploringTI}, and output speeds depend on individual screen reader settings \cite{bragg_listening_2021}. We approximate time and effort using user and system word counts, contextualized by conversational turns, submitted images, goal achievement, and scenario characteristics. 

\begin{figure*}[t]
    \centering
\includegraphics[width=0.8\textwidth,angle=0,clip=true,trim=0 0 0 0]{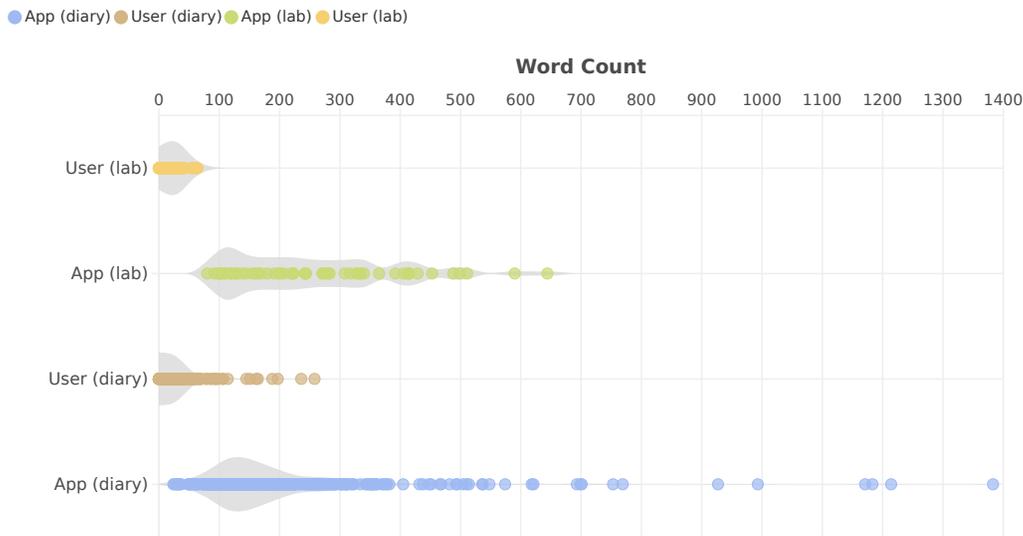}
    \caption{Lengths of user inputs and system responses (measured in word count) across in-lab session and diary session interactions.
}
    \Description[]{Violin chart showing the word count distribution of user inputs and system responses during in-lab and diary session interactions. In the in-lab sessions, user inputs are all under 100 words, while system responses typically range from approximate 80 to 500 words. In the diary sessions, user inputs mostly fall between 0 and 100 words, with a few outliers extending to 150–300 words. System responses during diary sessions show a broader range, primarily between 30 and 400 words, with some responses reaching up to nearly 1,400 words.}
    \label{fig:word-count}
\end{figure*}

\autoref{fig:word-count} shows a violin plot of interaction lengths for user inputs and system responses across both lab and diary sessions. The x-axis measures length in word counts. The width of each violin at a given count reflects the density of interactions---wider sections indicate more interactions at that length. The horizontal span shows the overall range of observed word counts. At a high level, responses are over ten times longer than user inputs.
In the lab, user inputs have a median length of $20$ words (range $0$-$64$). Diary interactions show a  lower median of $10$ words but a much wider range, extending up to $258$ words. 
This longer tail, likely reflects the nature of these real-world interactions, where participants had greater flexibility and autonomy in how and when they engage with the system---conditions that encouraged more detailed or context-rich input. 
The system is consistently more verbose than users in both settings.  Responses have a median length of $242$ words in the lab and $160$ in the diary. Although generally shorter in the diary, the distribution shows a longer tail, where some diary interactions prompted more elaborated or multi-turn conversations.

\begin{figure*}[t]
    \centering
\includegraphics[width=\textwidth]{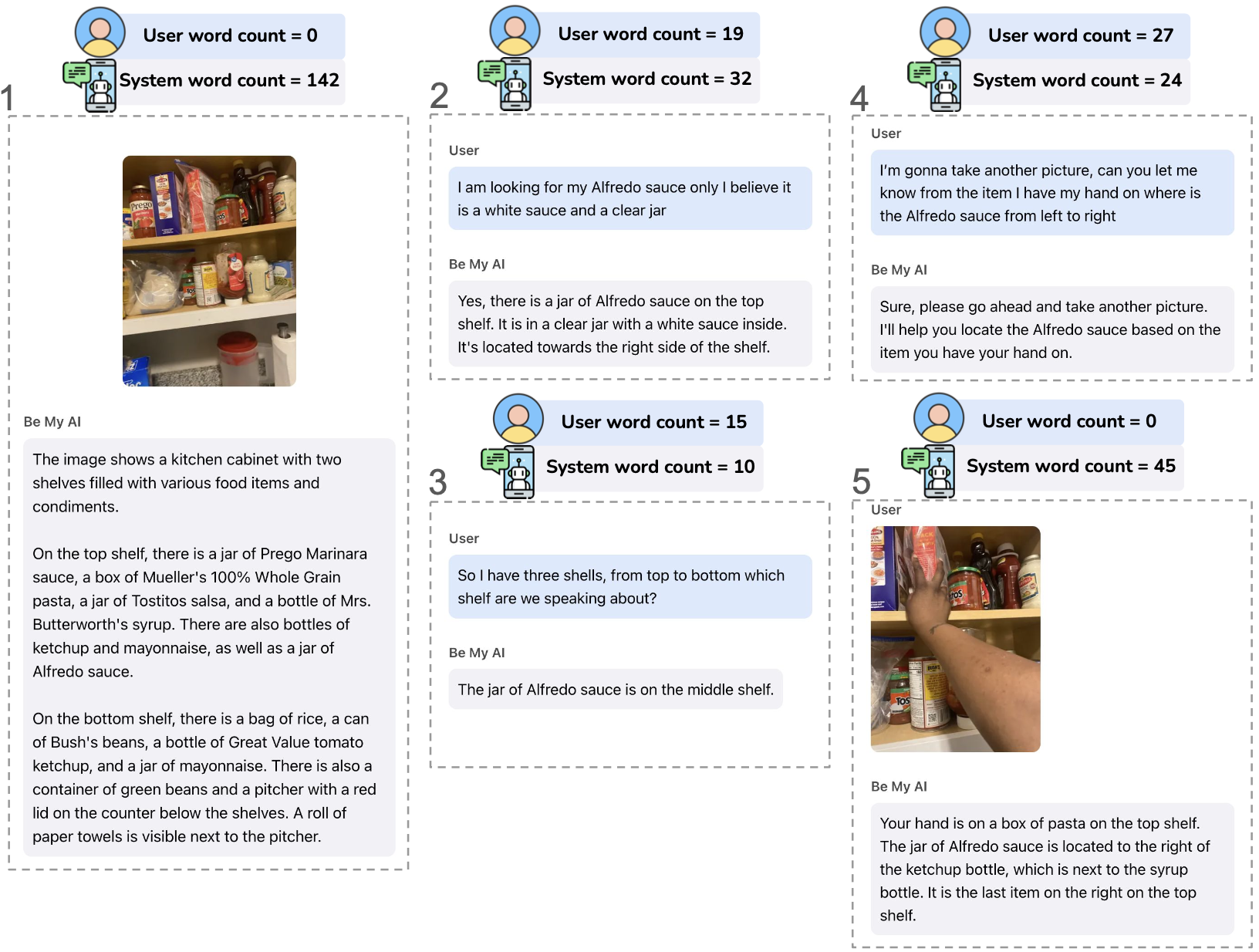}
    \caption{Example interaction consisting of five conversational turns with word count for each turn.
}
    \Description[]{an interaction example consisting of five turns between a blind user and the VQA system, including two images and three text prompts. The interaction shows a participant locating a box of snack in their cabinet. In the first turn, the user submits a image, and the system responds with a detailed description. The second turn to the fourth turn consist of text prompts from the user, followed by shorter responses from the system. In the fifth turn, the user shares another image, prompting a slightly longer response from the system. Participant input word count and system word count are variable in each turn.}
    \label{fig:interaction-ex}
\end{figure*}

\subsubsection{User Input Length in Context}
In the diary, $10\%$ of interactions ($n=38$) end with the system's initial response and no follow-up from users (input length is 0). Of these, only $3$ involved a second image upload after a low-quality first image. In such cases, the response had $85$ words on average ($SD = 37$). In nearly all instances, users reported that the initial response met their goal; the only exception was a low-quality image case, where the user chose not to continue.

In contrast, in $82\%$ of diary interactions ($n=302$), users followed up with inputs of $1$-$60$ words. These interactions averaged $4$ conversational turns ($1 - 21$, $SD=2$) with 1 to 4 images submitted.
In $60\%$ of cases, users submitted only one image, suggesting that the first image was often sufficient for a useful response.
System responses were longer in this group, averaging $207$ words ($SD = 150$). Most interactions ($83\%$) were considered successful by users.

Despite the overall success of most interactions, some were unsuccessful, typically falling into three categories: missing or inaccessible information, model limitations, and model errors. A common issue was that the needed information was absent from or unreadable in user images, often involving text on packaging. For example, P9 shared, \textit{``It is very difficult to use this app with bottles. I could not get a clear picture. I even asked the AI how I could improve the picture and tried many different angles and distances, and I still wasn't able to get a clear picture.''}
Submitting additional images resolved $66\%$ of such cases, though some questions remained unanswerable.
Other failures arose when requests exceeded the model’s capabilities. For instance, P6 asked about sofa dimensions to select a furniture cover---an inference the model is not capable of making. \revised{In fact, we noticed a persistent trend where the system avoided guessing measurements and distances in space (\textit{e.g.}, when P7 asked,\textit{``If you were to guess, how deep is the snow?''}).}  Several participants sought advice on medications, which the system is not qualified to provide. In these cases, the model appropriately directed users to professionals.
Finally, some interactions failed due to model errors. Users often recognized inaccuracies when responses contradicted prior knowledge or did not hold up in practice. For instance, when P10 asked for the expiration date on a bag of salad, the system responded with ``February 2024.'' P10 noted, \textit{``I knew it couldn't be that old.''}
In contrast, a few participants accepted incorrect outputs---typically in text-reading tasks involving complex layouts, such as microwave control panels or washing machine knobs---where verification was difficult and users defaulted to trust.

We also observe a long tail of interactions ($n=27$, $7\%$) with substantially longer user inputs ($60$- $258$ words). These typically involve more conversational turns, with a median of $6$ turns ($2-17$, $IQR= 3-12$),
reflecting deeper dialogue. Extended exchanges arose  when users made multiple attempts to capture usable images, required multiple clarifications, asked curiosity-driven follow-up questions, or requested updates on dynamic scenes.  
We further explore this iterative process in the next section.
Not surprisingly, these longer interactions included more images ($1$-$11$, median=$3$) and longer overall system responses ($36$-$1383$, median = $309$). 
Despite their complexity, $78\%$ were considered successful; the unsuccessful cases involved complex tasks combined with model errors, leading users to give up after extended attempts.

\subsubsection{\revised{Patterns in Conversational Turns}}

\revised{To examine structural patterns in how participants progressed through their interactions, we applied $k$-means clustering ($k=3$) using two features: the number of conversational rounds and whether  participants reported successfully completing the task. We selected $k$ based on interpretability. This analysis on the diary data, surfaced three distinct clusters that reflect common patterns of turn taking and reported outcomes.}

\revised{\textit{Cluster 1: Short-round interactions reported as successful (n = 286).} The majority of the interactions involved only a small number of conversational rounds (median = 2 rounds, range = 1–7) in which participants indicated that they obtained the information they needed to complete their task  with only light prompting. All participants had at least one interaction that fell into this cluster. An example of this appears in tasks related to identification and visual exploration. For instance, P8 captured a clear photo of the view outside an airplane window. The system correctly described the scenery, and after she responded with appreciation, it replied with \textit{``I’m glad I could help…''} (a total of 2 rounds). Reflecting on this interaction, she explained: \textit{``I got the response that I was looking for. I wanted to know what was going on outside of the window seat.''} }

\revised{\textit{Cluster 2: Short-to-medium-round interactions reported as unsuccessful (n = 53).}
These interactions consisted of a modest number of rounds (median = 3, range = 1–9) but were reported as not meeting participants’ needs. Participants rarely used any prompting and often stopped early, which may indicate that they encountered issues they felt unlikely to resolve within that interaction. Interactions in this cluster came from only about half of the participants (P1, P5, P6, P8, P9). We typically observed this type of interaction when the camera frame did not capture the information needed to interpret the scene. For example, P6 made 4 rounds with Be My AI while asking for directions on using a razor. Despite taking three images, she was unable to include the part of the box that containing the instructions and ultimately discontinued the interaction.}

\revised{\textit{Cluster 3: Long, effortful dialogues with mixed outcomes (n = 24).} These extended interactions (median = 10.5 rounds, range = 8–21) showed the highest use of customization techniques (median = 2 techniques, range = 0–8). Participants described these cases as requiring substantial troubleshooting or refinement of outputs, and outcomes varied. Interactions of this type came from six participants (P9, P6, P2, P11, P3, P10). These appeared to be moments where participants relied more heavily on prompting techniques to guide the system toward responses they considered more useful. They typically involved participants feeling encouraged to continue when the system’s reassuring or seemingly accurate responses signaled progress. For example, P2 used decomposition and grounding while searching for ranch dressing in her refrigerator, continuing for 10 rounds until she located it. As she noted, \textit{``I knew that the app was correct.''}}

\subsubsection{Scenario Characteristics as Context}
\begin{figure*}[h]
    \centering
    \includegraphics[width=\textwidth]{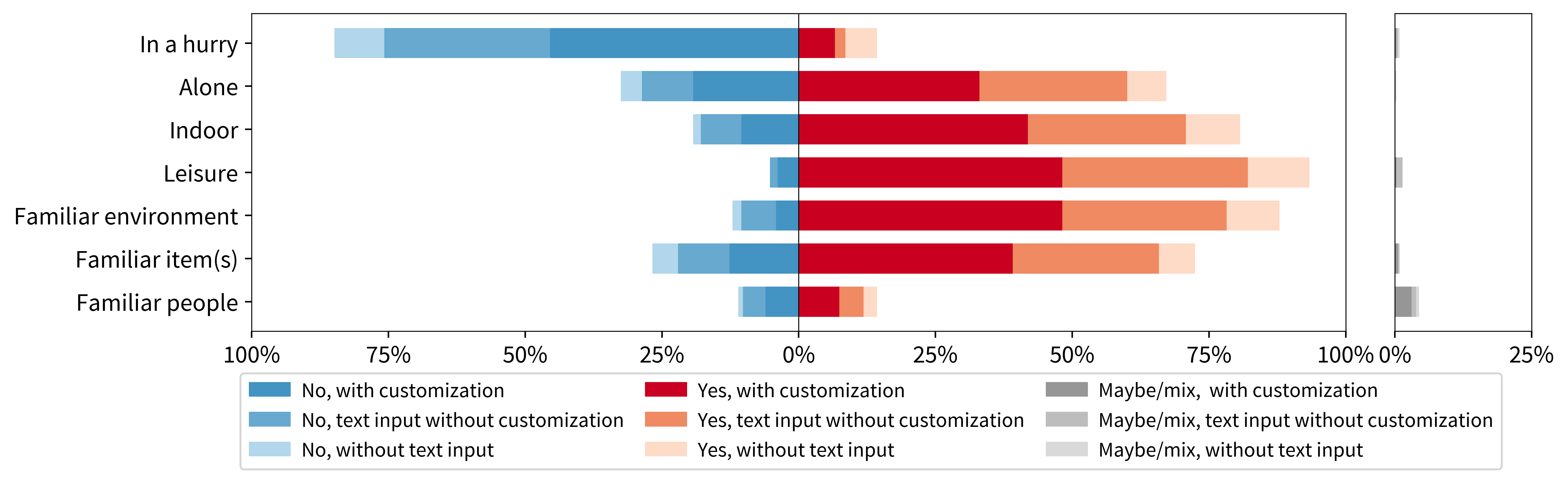}
    \caption{\textcolor{black}{
    Distribution of interactions across scenarios, with colors indicating presence, absence, or uncertainty and intensity marking customization. Proportionally, interactions involved more customization in familiar environments and with familiar items, slightly more around others than alone, similar indoors and outdoors, and less when in hurry.
    }}
    \Description[]{Stacked horizontal bar chart illustrating preferences for interaction types across various contextual factors, categorized by levels of customization and input types. The y-axis lists seven situational factors: "In a hurry," "Alone," "Indoor," "Leisure," "Familiar environment," "Familiar item(s)," and "Familiar people." Each bar is segmented into multiple color-coded responses indicating different combinations of customization and text input: dark blue (No, with customization), medium blue (No, text input without customization), light blue (No, without text input), dark red (Yes, with customization), orange (Yes, text input without customization), peach (Yes, without text input), and three shades of gray for "Maybe/mix" options. The x-axis is percentage-based from 0\% to 100\%, showing distribution of responses. The right side features mirrored bars for "Maybe/mix" responses. The chart reveals that participants are least likely to engage when "in a hurry" (high "No" responses) and most likely to engage when in "familiar environments" or during "leisure" (high "Yes" responses with customization or text input). A legend at the bottom provides detailed color definitions for interaction modes.}
    \label{fig:scenario_coding}
\end{figure*}

Based on participants’ text descriptions of their contexts and goals, we annotated the scenarios in their interactions and verified these annotations in post-interviews.
\revised{The bar graph in \autoref{fig:scenario_coding} shows the distribution of all interactions across different scenario characteristics (y-axis). Most cases reflect a clear presence (red) or absence (blue) of the characteristic, with relatively few uncertain cases (gray). Bar lengths indicate the percentage of interactions in each scenario. Most logged interactions occurred during leisure activities ($93.4\%$) and when participants reported being not in a hurry} ($84.8\%$), often in familiar environments ($87.9\%$) or indoors ($80.7\%$). Social context \revised{was also associated} with when interactions took place: most occurred when participants were alone ($67.2\%$) with fewer interactions happening around familiar people ($17.1\%$), strangers ($11.0\%$), or mixed groups ($4.4\%$). Similarly, most interactions involved familiar items ($72.5\%$).
Overall, participants used the conversational VQA system  mainly in everyday, comfortable routines rather than novel or high-stress situations. It remains unclear how lengthy or erroneous system responses influence these choices.

\revised{The color-intensity segments of each bar in \autoref{fig:scenario_coding} show the proportion of interactions that involved user text input and customization under each scenario.
At a high level, we observe that} participants tried customization techniques more often in familiar environments ($55\%$) than in unfamiliar ones ($34\%$). A similar pattern emerged with items: $54\%$ of interactions with familiar items involve customization versus $47\%$ with unfamiliar items. Social context also mattered but in a different way: participants customized most when surrounded by familiar people ($60\%$), slightly less with strangers ($56\%$), and least when alone ($49\%$). Physical setting had little effect, with similar rates indoors ($52\%$) and outdoors ($54\%$). Customization occurred in $46\%$ of hurried interactions, compared to $54\%$ when participants were not in a hurry. In professional contexts, participants customized far more ($74\%$) than in leisure settings ($52\%$), an unexpected finding given that casual environments might seem more conducive to exploration.





\subsection{Participants' Tactics for Tailoring Responses}\label{sec:4.2}





\begin{figure*}[h]
    \centering
    \includegraphics[width=\textwidth]{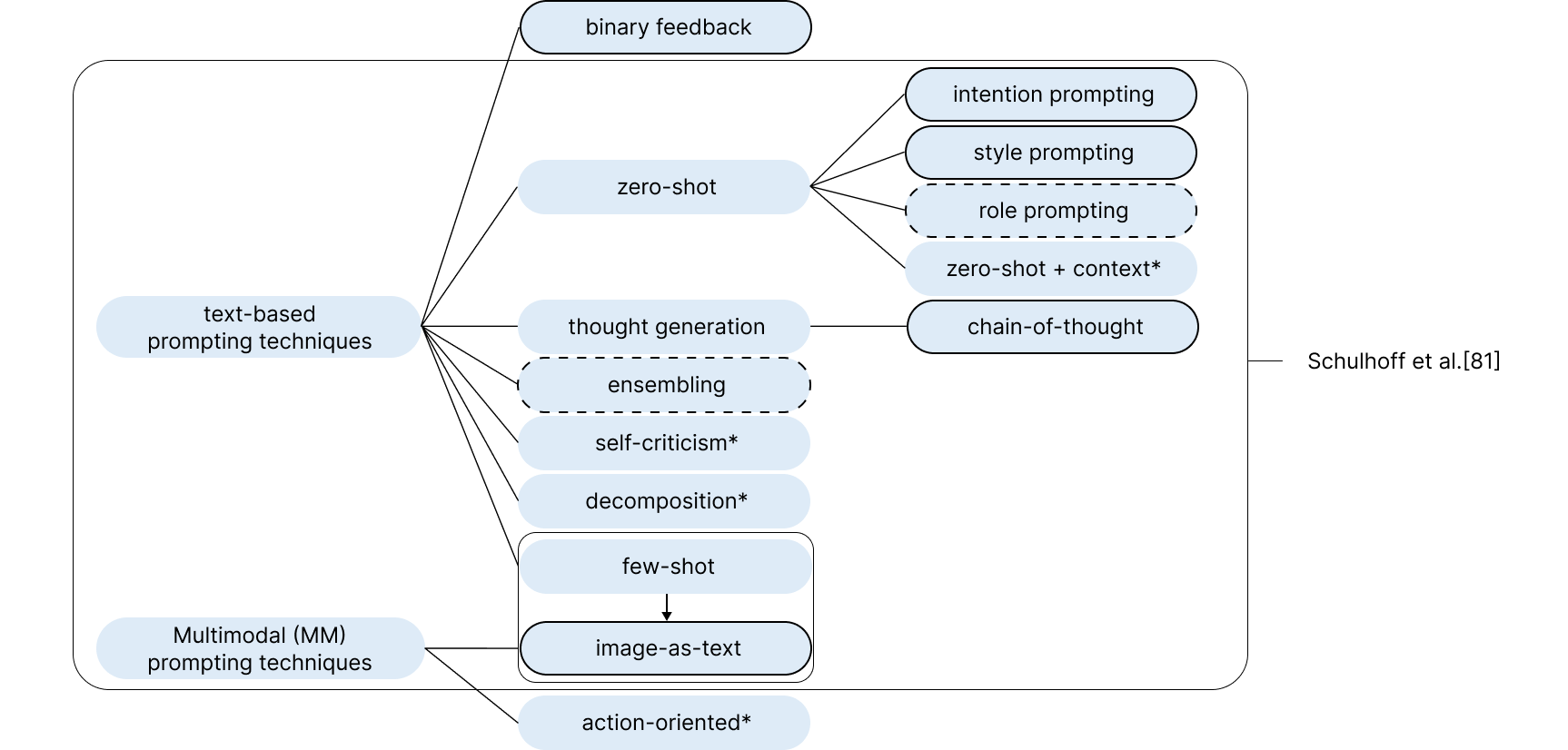}
    \caption{All the customization techniques observed in \datasetname~dataset following the hierarchy from \cite{schulhoff2025promptreportsystematicsurvey}. The techniques with solid border are introduced in the lab. Techniques with asterisk come up by participants in the diaries. Techniques with dash border are customization techniques that participants' explored strategies resemble. Due to the limitation of \textit{Be My AI}, we use image-as-text prompting instead of few-shot prompting to provide information.}
    \Description[]{Diagram showing different types of AI prompting techniques from Schulhoff et al. It splits into two main categories: text-based prompting and multimodal (MM) prompting. Under text-based prompting, several methods are listed: Binary feedback, Zero-shot, which branches into intention prompting, style prompting, role prompting, and zero-shot with context*, Thought generation, leading to chain-of-thought prompting, Ensembling, Self-criticism**, Decomposition**, Few-shot, which connects to image-as-text prompting. The multimodal prompting branch includes image-as-text and action-oriented* techniques. Some methods are marked with an asterisk to highlight variations. The chart emphasizes how different prompting \revised{techniques} connect and build on each other.}
    \label{fig:taxonomy}
\end{figure*}

In this section, we examine how blind users interact with the VQA system, focusing on how they attempt to control and tailor responses to their needs. Our analysis of the interactions--both in the lab and participants' diaries--along with their reflections, reveals several patterns.
By default, the system provides an image description with each input, but participants often find these descriptions lengthy, broad, or insufficient for completing a specific tasks. 
The most common strategy is to ask direct questions without tailoring. Beyond this, participants carry over prompting techniques from the lab and experiment with new ones, as shown in Figure~\ref{fig:taxonomy}. 
Through these approaches, they sought to control verbosity, direct attention, and better achieve their goals.

\subsubsection{No Prompting Technique}
\label{subsubsec:no_prompt_tech}

We observe that the most prevalent way blind users interact with the system is by simply asking questions without explicitly attempting to further tailor the answers. This is the case even in the lab, where participants were instructed to prompt and customize. \revised{In some cases, participants mentioned that customization techniques did not always feel intuitive and that there was a learning curve. As P3 explained: \textit{``I don't think it will come natural. I mean, I even forgot about that aspect of it because when I'm trying to get information, usually the way to go is to ask for it. But I'm not close-minded to that personalized technique. .. I would have to get more familiar with it to use it.''}}

\revised{On average, $77.27\%$ of each participant’s in-lab interactions included at least one question ($50\%$–$100\%$, $SD = 20.78\%$).} In their diaries, where they chose use-case scenarios of their own, participants still asked a high proportion of questions, with at least one question appearing in $74.8\%$ of their interactions on average ($0\%-100\%$, $SD=29.81\%$). Both in the lab and in their diaries, we observe that with their questions, all ($n=11$) participants shape the responses to be more task oriented. For instance, in the first in-lab scenario of locating the door, participants are instructed to provide binary feedback related to the helpfulness of the description in the response. Yet, P2 asked \textit{``Where is the exit sign?''} after the first turn of interaction, aiming to narrow the response to help determine the general direction of the exit door. 

Indeed, in its current form, \textit{Be My AI} defaults in image description without a context for a task. Users are not allowed to submit a question alongside the first image, but through subsequent interactions.  More than half ($n=6$) of the participants found this limitation to be critical as it often forced them to sit through lengthy descriptions before getting to the information they actually needed. Reflecting on this after trying to locate her keys in an unfamiliar room, P6 shared: \textit{``when I took a picture, it gave me a bunch of information that I did not want. So I have a question [in mind] before I take a picture.''} Even when the goal was to obtain a description, in many of their interactions ($40.38\%$) participants follow up the default initial description with questions for more. For instance, P3 asked \textit{``What color are the pillows on the couch?''} for detailed descriptions after knowing there were several decorative pillows on the sofa from \textit{Be My AI}'s default description. Participants also ask \textit{Be My AI} to read specific text in $20.33\%$ of their interactions, often supplementing with additional images. For example, P10 asked \textit{``What is the expiration date''} of a yogurt with several images to let \textit{Be My AI} find and read the date.

\begin{figure*}[h]
    \centering
    \includegraphics[width=\textwidth]{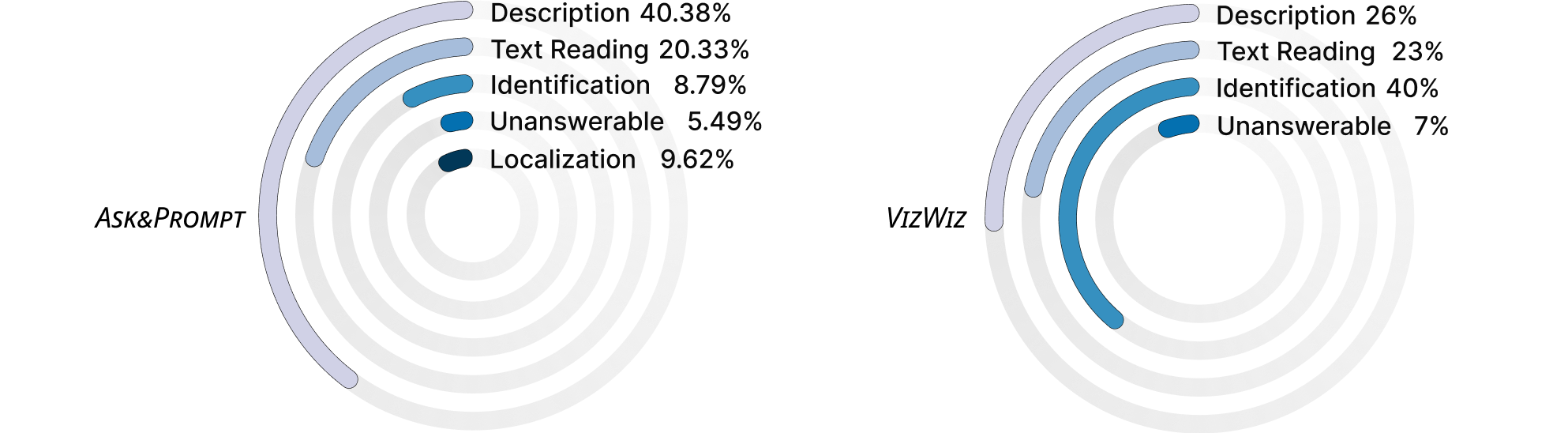}
    \caption{Distribution of question categories in 363 multi-turn interactions from participants in our diary study session and 495 single-turn interactions randomly sampled in VizWiz~\cite{Brady2013investigating}.}
    \Description[]{In participants' dairies, there are $40.38\%$ of interactions involve questions asking for description, $20.33\%$ of interactions involve questions asking for text reading, $9.62\%$ of the interactions involve questions for object localization and orientation, $8.79\%$ of interactions involve questions for object identification, and $5.49\%$ of interactions where the question remained unanswerable despite follow-up attempts.}
    \label{fig:question-category}
\end{figure*}

As shown in Fig~\ref{fig:question-category}, $9.62\%$ of the interactions in the diaries involve questions for object localization and orientation, while $8.79\%$ of interactions involve questions for object identification. Interestingly, with similar breakdown by Brady et al.~\citet{Brady2013investigating}, our distribution notably differs from that of VizWiz. The large drop of identification questions from $44\%$ in VizWiz to $8.79\%$ in our study could be explained by \textit{Be My AI}'s default image description right after the image, preempting the need for identification questions. We also suspect that \textit{Be My AI}'s support for real-time and multi-turn question answering explains the lower number of unanswerable questions from $7\%$ to $5.49\%$, with an additional $10.44\%$ of unanswerable cases in the diaries being resolved through the submission of additional photos. More so, we observe the emergence of a new question category pertaining to localization despite \textit{Be My AI} terms of service indicating that it is ``not designed or intended to be used as a mobility aid''~\cite{Bemyeyes2023terms}. 

Unsurprisingly, even though the purpose of the study was to investigate customization, asking questions was deemed a more natural and intuitive way for the blind participants in our study to interact with this VQA system. P7 preferred asking question to framing them with prompting techniques because asking question is \textit{``just more natural ... like having a conversation.''} P3, who did not use any technique but asked questions in $56.75\%$ of her diaries, concluded her way of interacting similar to \textit{``googling something''} because she \textit{``always ask a question''} to get the information. Although she acknowledged that, compare to asking questions, prompting techniques could, in theory, make responses more concise or relevant, she did not think they would come naturally. Yet, she still stayed open to the idea of using prompting techniques in the future and said, \textit{``I would have to get more familiar with it (prompting techniques) to use it... I hasn't transitioned [my way of interacting] in my mind yet with artificial intelligence.''}


\subsubsection{Binary feedback}
This is the first customization technique introduced in the lab situated within an exit door navigation task, where participants could provide one-word feedback on individual responses during the interaction. \footnote{\textit{Be My AI} only supports feedback through a ``thumbs-up'' or ``thumbs-down'' feedback button at the end of the chat but not after each individual interaction response unlike other commercial interfaces, e.g., ChatGPT, and thus this feedback mechanism does not directly impact the user’s ability to achieve their immediate goals.}

\revised{Binary feedback is a common interaction pattern across many interfaces and does not necessarily imply system adaptation. We introduced it as an accessible baseline to anchor discussions about feedback mechanisms and to contrast it with more expressive strategies. Its simplicity helped frame participants’ expectations about what “feedback” could mean, even within a system without memory.}

Almost all ($n=10$) participants practiced this technique after our lab introduction. Surprisingly, although many ($n=8$) of them gave \textit{``helpful/great''} feedback, some of them did not find the description they received sufficiently helpful.
For example, P3 gave a \textit{``helpful''} feedback, even though she did not manage to locate the door using the app's responses alone. One possible reason for this is that participants may have wanted to maintain a positive attitude toward the technology they were testing \cite{wixom2005theoretical}. As being the easiest prompting technique, however, more than half ($n=6$) of the participants did not prefer using it immediately in the lab. One key reason is its limitation of lack of specificity. For example, P5 said, \textit{``if I just say helpful, okay, helpful on what? I don't express, it's like nothing.''}

In their diaries, we observe that about half ($n=6$) of the participants made limited use of this technique, though distinct usage patterns still emerged. P7 indicated that binary feedback is more natural to be provided at the end, which is similar to what \textit{Be My AI} has implemented by the ``thumbs-up'' or ``thumbs-down'' feedback button at the end of the chat.
With greater flexibility during the diary study, half ($n=5$) of the participants expressed binary feedback at the end of the interactions.

\begin{figure*}[t]
    \centering
    \includegraphics[width=0.7\textwidth]{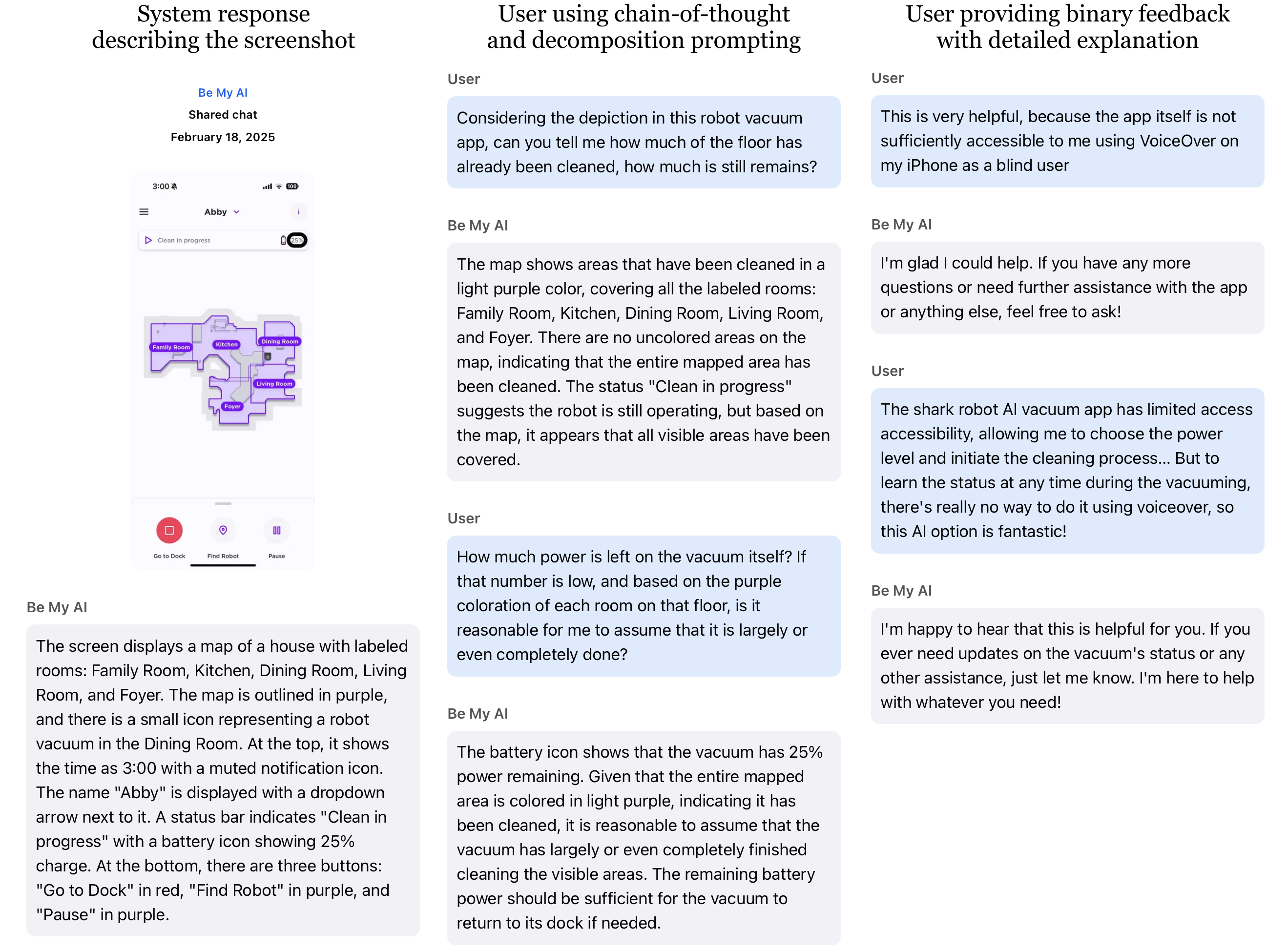}
    \caption{P11 using (a)) binary feedback along with detailed explanation in the style of chatting, (b) chain-of-thought prompting during the diary session, and (c) explicitly mentioning vision level.}
    \Description[]{An interaction example using three customization techniques. 1) Binary feedback along with detailed reasons in the style of chatting with the system: participant prompted the system with ``This is very helpful, because the app itself is not sufficiently accessible to me using VoiceOver on my iPhone as a blind user.'' 2) Chain-of-thought and decomposition prompting: participant broke down his goal of assessing vacuuming progress into multiple prompts. First, he asked for factual details such as the area cleaned and the battery level. Then he followed up with an inference question, ``If that number (the battery level) is low, and based on the purple coloration (cleaned area) of each room on that floor, is it reasonable for me to assume that it is largely or even completely done?'' 3) explicitly mentioned blindness: in the same prompt as binary feedback one.}
    \label{fig:interaction-binary}
\end{figure*}

Another observed pattern is that, in their diaries, participants naturally paired binary feedback with a short rational, addressing the lack of specificity they had noticed in one-word binary feedback during the lab study. This feedback style emerged organically through casual conversation with the system rather than as a deliberate strategy.
For example, P8 spontaneously shared her excitement with \textit{Be My AI} after discovering details outside her airplane window: \textit{``“Wow, this is my first time scanning a picture and realizing that there’s a plane parked next to the plane that I’m currently sitting in. I appreciate all the details.''} Similarly, as shown in \autoref{fig:interaction-binary}, P11 gave a \textit{``very helpful''} feedback along with the detailed reason. He further explained in the post interview that this kind of feedback is not always intentional and strategically planned: \textit{``It (the system) would just evoke a natural response.''} He further explained that a more human-like voice, such as the one used in PiccyBot~\cite{piccybot}, could further encourage this behavior: \textit{``It’s literally like you’re talking to a person. So it almost invites you to respond, even if you’re annoyed with it, or if it did a really good job, saying wow, that’s awesome, that’s so helpful.''}

\subsubsection{Zero-shot: style prompting}\label{sec:4.2.3}


This is the second customization technique that participants are introduced into within the exit door navigation task. About half ($n=6$) of the participants tried it in the lab following our prompting examples using \textit{``less details''} or \textit{``more details''} to \textbf{control the verbosity} of responses. Some ($n=3$) went further by adding additional questions or instructions guiding the system to filter the information for them. For example, P8 prompted with, \textit{``Less details,.. just tell me what is very close to me''}.

In their diaries, we observe a similar usage pattern. More than half of the participants ($n = 6$) used style prompting paired with specific instructions or other techniques to control the verbosity of the responses. They primarily used it to receive shorter, more focused answers. For example, P10 repeatedly framed prompts as \textit{``Tell me only about ... and nothing else around it''} using this pattern each time to filter for relevant information. Indeed some participants ($n=4$) explicitly reported that they would use style prompting as a filter to \textit{``eliminate the unnecessary stuff''}. P2 described this approach as intuitive, likening it to natural conversation: \textit{``that's just like having a conversation with somebody, and they're going on and on and on, they're long-winded.''}
P3, who had prior experience with \textit{Be My AI} for months, immediately found style prompting useful in the lab because it finally addressed the issue that \textit{Be My AI} \textit{``gives way too much information sometimes.''} However, she did not use this technique in her diaries. We suspect that the reason is still her unfamiliarity to techniques mentioned in ~\ref{subsubsec:no_prompt_tech}. 



\subsubsection{Zero-shot: intention prompting}
\revised{We introduced intention prompting during the nutrition-fact identification task in the lab, clarifying that it can reduce repetition and streamline interactions, particularly when users have multiple queries tied to the same underlying goal (e.g., dietary assessment, cooking progress, etc.).} Although most ($n=9$) participants practiced it in the lab, some initially struggled because they were accustomed to phrasing their intentions as direct factual questions.
For example, P6 began with, \textit{``It’s important for me to eat lower carb and high protein, how many grams of protein is this?''} After further guidance, she refined the prompt to convey her broader goal and invite an open-ended response: \textit{``I follow a low carb diet. Is this a good food for me?''}
In their diaries, we observe that this habit of  interpreting intentions as concrete questions persisted. For example, P8 framed her intention of deciding whether to wear high heels to a concert as a concrete question asking whether the outside was wet. Interestingly, she independently noticed this during the post interview and proactively improved her prompting style, explaining: \textit{``when we are more intentional in what we're asking, then we're going to get more specific responses.''}


Almost all ($n=8$) participants who tried this technique in the lab said they would use it again and even envisioned future use-case scenarios. For example, P5 imagined applying it while grocery shopping: \textit{``I have more details and some advice... that can make me start to make my grocery by myself.''} Similarly, P3 pictured using it to choose a lipstick color independently.
Indeed, in their diaries, one of the main uses of intention prompting was for requesting suggestions. For instance, P6 shared a photo of her kitchen and prompted directly: \textit{``I want to remodel my kitchen, I need suggestions''}.

In the post interview, many participants ($n=8$) mentioned their preference towards intention prompting. For example, P7 explained that this technique helped \textit{Be My AI}  \textit{``understand where I'm trying to go, as opposed to me having to direct it [pause] exactly the specific point.''}
However, initial reactions varied. In the lab, P10 and P11 both expressed reluctance to use intention prompting, expressing confidence in their own questioning skills and limited trust in the system’s ability to infer intent. P10, focused on reducing sodium intake, said: \textit{``I don't think I would tell the app that (my intention), I would just ask how much sodium is in here... cause I don't think it's necessary to get the answer that I want.''} Interestingly, both later adopted intention prompting repeatedly in their diaries and named it a favorite technique in the post interview. P10 described it as feeling \textit{``natural''} to use while P11 appreciated its efficiency: \textit{``because you can certainly narrow the lane dramatically [by expressing the intention] ... and you'll get to the finish line much more quickly.''} P11 also discovered that \textit{Be My AI} could carry his stated intent across follow-up images, calling this \textit{``a big time saver which I hadn't appreciated.'' }

\subsubsection{Chain-of-thought (CoT) and decomposition prompting}
CoT prompting is introduced to participants in the lab within the item localization task by encouraging them to include reasoning behind their queries. We observe that this technique appears to be the most challenging for participants to grasp. In the end, only about half of the participants ($n=5$) attempt to use it in their prompts. For example, P8 incorporated reasoning by prompting \textit{``I want to know if there are any cups with liquids near the Lipton soup because I don’t want to spill anything.''} Even after researcher's explanation and examples, some participants ($n=4$), still defaulted to either style prompting, intention prompting, or simply asking questions. Use of CoT remained limited in their diaries as well. One notable exception was P11, who repeatedly applied CoT when interpreting screenshots. As shown in Fig~\ref{fig:interaction-ex}, he guided \textit{Be My AI} to integrate multiple visual cues (e.g., color coding, battery level, room labels) to infer a vacuum’s cleaning status.
We speculate that his familiarity with CoT may be attributed to his extensive experience interacting with \textit{Be My AI} and other GenAI systems. 

\begin{figure*}[]
    \centering
    \includegraphics[width=0.5\textwidth]{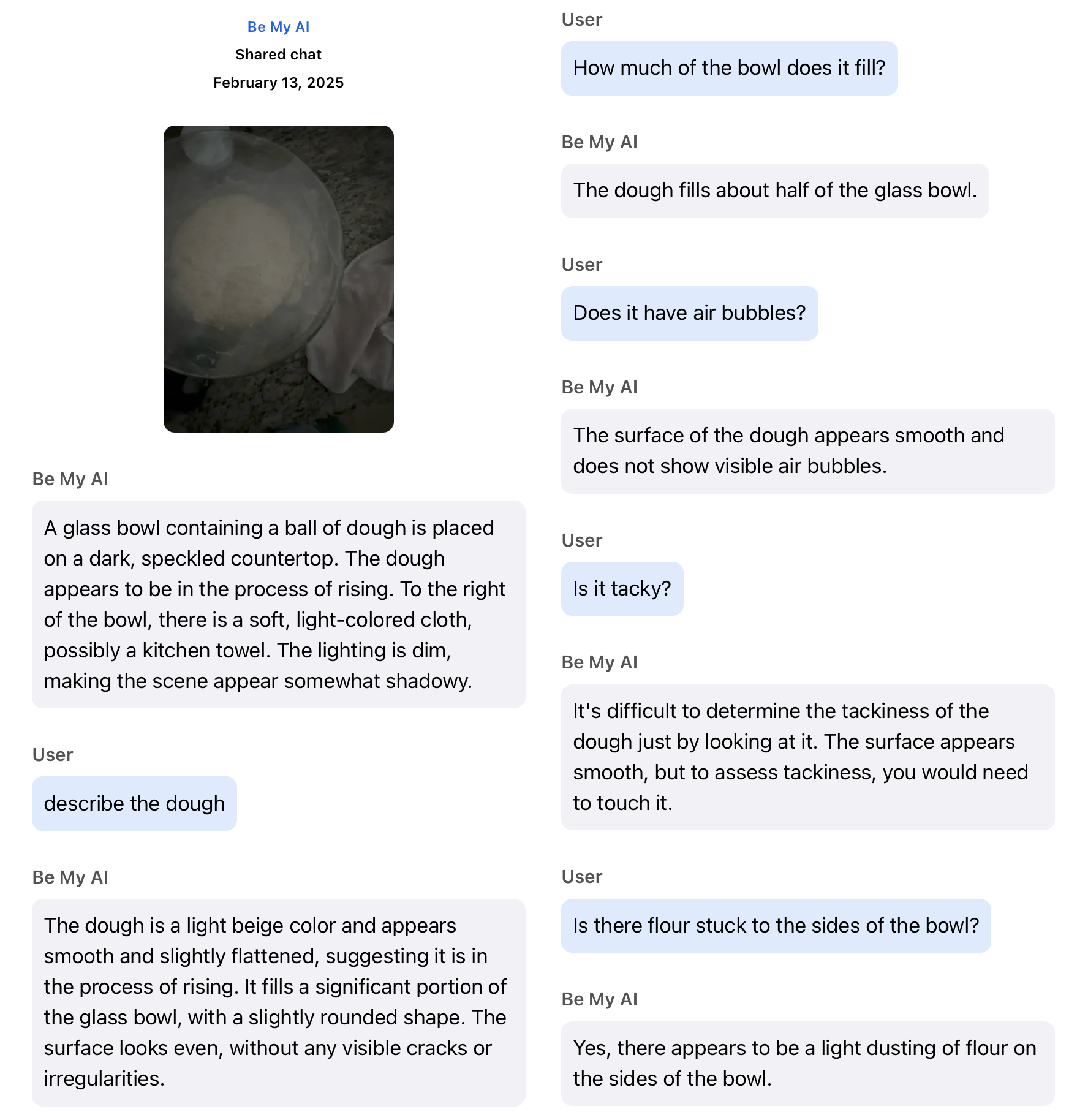}
    \caption{P9 using decomposition prompting in the diary.}
    \Description[]{An interaction example using decomposition prompting. Participant breaks down her overall goal into a series of questions, asking the system to check its volume, look for air bubbles, assess whether it is tacky, and determine if any flour remains in the bowl.}
    \label{fig:interaction-cot-dough}
\end{figure*}

Although we introduced CoT in the lab, three participants showed at least one instance of interpreting it as decomposition prompting, breaking a task into subtasks, or a series of questions to guide the system. For example, P9 decomposed the item localization task into two parts by prompting \textit{``I'd like to know where the soup is and what objects are around it.''} This technique is even more common in their diaries, where more complex real-world scenarios occurred. Many ($n=9$) participants decomposed tasks to get more precise, step-by-step responses. For instance, P2 decomposed her task of locating a food box on a shelf by asking:\textit{``From top to bottom which shelf is it located on, and from left to right exactly where is it located? Is it visible or behind something?''} P9 also applied this technique in cooking scenarios. As shown in \autoref{fig:interaction-cot-dough}, she breaks down the assessment of dough readiness into several descriptive questions to gather detailed information.



Many participants ($n=8$) expressed a clear preference towards decomposition prompting during the post interview. P7, who used decomposition prompting in the majority of his diaries,
emphasized that this technique was particularly useful when dealing with deviations, instances when responses began to drift from the goal he had in mind. He explained that breaking the task into smaller steps let him pull the conversation back on track.
In addition to this, participants valued decomposition prompting as a way to obtain more complete and accurate information. P9 noted that although it can be time-consuming, breaking down the task is \textit{``the only way to make sure that it (the response) was 100\% correct.''} Similarly, P11 appreciated that decomposition prompting enabled him to get granular details without relying on the system’s own filtering, especially for image description. He adopted this technique after the AI once omitted elements it judged \textit{``inappropriate,''} rather than describing everything visible \textit{``as objectively and neutrally as possible.''}
As many participants independently came up with this technique in their diaries, they later reflected on its naturalness in the post interview. For example, P9 found her interactions with \textit{Be My AI} similar to those she had with sighted volunteers unfamiliar with the context, which led her to apply the same step-by-step prompting she typically used with people to her interactions with the system.


\subsubsection{Image-as-text prompting}

This is the last customization technique that participants are exposed to in the lab, situated in the personal item localization task. 
Similar to teachable object recognizers~\cite{kacorri2017people, findmything}, participants envisioned that it will be useful to locate personal items or distinguish between similar objects. For instance, P10 imagined to identify her cane at a large conference and said, \textit{``If you ever been to a conference where there are 5,000 blind people, and you want to know which cane is yours. In that case, I will put something on my cane, like some kind of keychain or charm, so that if they do get mixed up I know which one is mine. So that's a question that I could then ask.''}

In their diaries, we observe similar use patterns to distinguish objects. For instance, P11 used this technique to identify a sanitizer bottle next to a similar one by referencing a known visual detail: \textit{``The sanitizer has the blue cap.''}
Participants also used this technique to shape the focus of the interaction. For example, P8 prompted, \textit{``I am the woman in the blue and white striped dress holding a white cane,''} to narrow down the focus to herself. She then added more specific details, e.g., \textit{``coconut booties''} and \textit{``the silver Tiffany stud earrings in both my ears''}, to check each item she was wearing, ensuring that she looked good in the photo.

Almost all participants ($n=10$) expressed a preference for this technique and anticipated using it in the future when asked in the lab. Yet, this technique appeared in a very limited number of diaries ($6.61\%$). We theorize this contrast is explained by the need for one to be able to recall detailed visual characteristics of objects or scenes for them to be described in detail to \textit{Be My AI}.
For instance, participants often looked for or located items they could typically identify by name or recall a few key properties. For example, when P2 asked \textit{Be My AI} to look for a Pop-Tart on the shelf, she was uncertain, \textit{``maybe strawberry flavor.''} In the post-interview, she expressed a willingness to improve her prompt using image-as-text prompting, yet still struggled with not knowing the appearance and having nothing specific to describe.
This difficulty even arise with personal items that participants use everyday. For instance, P3, the only participant who did not practice this technique in the lab, felt uncertain about the appearance of her keychain and found it hard to describe. She said, \textit{``I don’t think I would naturally do that. Oftentimes, maybe I don’t know the details of it.''} We believe that few-shot prompting, the original technique we wanted to introduce and not supported by \textit{Be My AI} could help overcome unfamiliarity with object and scene appearance even for familiar objects that may be the key reason for the limited use of this technique.

\subsubsection{Customization techniques beyond the ones introduced in the lab}

Beyond the customization techniques introduced in the lab, we observed that participants spontaneously combined zero-shot prompting with rich contextual information as a form of grounding.
Regarding the reason of providing the rich context, P2 said, \textit{``The more feedback or content you give it, the more feedback and content that it will give you.''} She concluded it as a way to prompt the system to generate more accurate and relevant responses. In many cases, the contextual information participants provide not only supports the current task but also implicitly signals personal preferences or patterns that could improve alignment over time.

In the diaries, participants also encountered situations where \textit{Be My AI} provided responses that were clearly incorrect. In these cases, participants naturally employed self-criticism prompting to address sycophantic behavior by questioning and correcting the original response\cite{schulhoff2025promptreportsystematicsurvey}. 
For instance, after \textit{Be My AI} responded with a price of $\$699.90$ for a dog bed in a supermarket, P7 questioned the system by asking, \textit{``Does the price actually say $\$69.99$?''} In some cases, participants followed up by asking the system to generate the response again after issuing a correction. We observed that \textit{Be My AI} provided correct information after participants used self-criticism prompting technique. However, it remains unclear whether the system truly revisited the image to verify the correction. Participants still face the risk that the system simply overagree with their correction rather than independently validating it.



Participants also explored prompting \revised{techniques} that resembled widely used customization techniques. One such strategy involved explicitly stating their blind identity in the prompt. \revised{In fact, neither researchers nor participants know whether Be My AI uses a custom system prompt or dataset; Surprisingly, we observed cases where explicitly revealing blindness changed system behavior.
}For example, P9 prompted the system to shift the reference point from the image framing to surrounding objects by saying, \textit{``I need to know what objects are around it because I’m blind and can’t see what is framed in the photo.''} \revised{In response, the system provided more comprehensive descriptions of nearby items and their relation to the object of interest.}
Although participants did not explicitly assign a role to the system, their use aligns with role prompting~\cite{schulhoff2025promptreportsystematicsurvey}. Thus, we suggest that role prompting can be a possible strategy for blind users to better align the system’s responses with users’ needs and expectations.

In addition to strategies resembling commonly used customization techniques, participants came up with a new prompting \revised{technique}---including their action in the prompt. Typically, they prompted the system with the action they were about to take in the next image.
For example, P2 prompted the system before uploading an image with her hand in by \textit{``I’m gonna take a picture of me touching a bottle, from that picture. Please let me know if I need to go left right up or down.''}
This action-oriented customization strategy illustrates how blind participants use prompts to shift the system’s focus in alignment with the evolving steps of a real-world task. This also shows the potential of VQA systems to support ongoing, real-time activities through conversational interaction.

\subsection{Other Opportunities for User Control and 
Customization}\label{sec:4.3}

In the previous section, we saw the customization techniques that participants used; some of them they were exposed to and some of them they were able to come up with. In this section, we identify other opportunities for customization that while might have not appeared in participants' interactions they could be useful based on participants feedback on 
the ideal response that they would expect.
We discuss these in the context of the tasks that the participants were trying to perform since ideal responses among tasks shared common themes.




\subsubsection{\textcolor{black}{Localization tasks call for verbosity control and personalization of spatial points of reference and safety cues}}

In the lab, participants consistently valued detailed descriptions when navigating unfamiliar environments, as these helped them build a clear mental map of their surroundings. As P4 put it when locating the exit door in the lab: \textit{``I like more information to know what’s around me.''}
However, in their diary interactions, where localization tasks were often performed under time pressure, preferences shifted. In these hurried contexts, participants most appreciated concise, task-focused responses and expressed frustration when the system added information irrelevant to the immediate goal. For instance, P1, when searching for her TV remote at home, explained: \textit{``If it was able to just say…we don’t see the remote, that would have been helpful so I could take the pictures faster without having to cipher through everything.''} This contrast strengthens our previous observation (Section~\ref{sec:4.2.3}) on the need for \textbf{verbosity control}, where users can toggle between detailed or concise output depending on context.  

A second theme concerned \textbf{points of reference} in spatial descriptions which was found both in the lab and diary interactions. Some participants preferred object locations described relative to themselves or relative to their hand when pointing, aligning with prior findings that blind users commonly include their hand, either touching or placing it near the target, as a reference during localization tasks~\cite{lee2019hands}. For example P2 wanted responses like \textit{``to the left of your hand''} 
often paired with distance cues such as \textit{``... from where you was just positioned, go to maybe the left, or… take five steps''}. Others preferred locations described relative to familiar landmarks. For example,  P11 said, \textit{``If you knew where something was, like the laptop… you could use it as a fixed point of reference''}. These preferences contrasted with \textit{Be My AI}’s defaults that were relative to the image frame (e.g., \textit{``top left of the image''}), which participants found less accessible and thus less actionable. 

Participants often expressed a desire for \textbf{safety awareness} beyond the target object with information about nearby or potentially hazardous items to support safe navigation. As P8 explained, \textit{``I hate the fact that… sometimes we just reach out into space… people reach and they knock you in the head, they’ve spilled stuff on you.''} This emphasis on environmental caution echoes prior findings \cite{Gonzalez2024Investigating}, who identified the avoidance of dangerous or unwanted objects as a frequent use case for blind participants using AI scene description tools.

\subsubsection{\textcolor{black}{Description tasks call for detail-level control and personalization on attributes, order, and time-lapse}}
These tasks involved requests for information about objects, scenes, screenshots, or people. In these cases, participants typically valued factual and relevant information. We also saw opportunities for customization in defining \textbf{attributes of interest}. For example, P11 wanted fine-grained color distinctions when asking about clothing: \textit{``It’s particularly helpful if it tells me the shade of the color … there must be dozens of shades of blue that can be discerned, and they can be quite different.''} By contrast, P10 focused on broader color categories (e.g., black, red). Prior work suggests detail preferences may relate to vision history~\cite{stangl2020person}: P11 lost sight later in life, while P10 was blind from birth. Such differences highlight potential for system-level customization through \textbf{memory of user preferences}~\cite{openai_memory_2024}, enabling responses tailored across sessions.

Echoing prior findings~\cite{morris2018rich}, participants also expressed preferences for the \textbf{order of information}. For example, P6 preferred cabinet descriptions to begin with style and color before number and layout. Beyond verbosity control, this points to the need for tailoring what kinds of details appear at each level. When no fixed preferences existed, participants favored a \textbf{hierarchical response style}: starting with a high-level overview, then offering ways to drill down. Dense outputs could be overwhelming (P7: \textit{``If it’s just a lot of information, I can’t process at all''}), but participants appreciated when the system inferred intention and offered follow-ups (P11: \textit{``I would have liked the overall perspective first and then have it go on its own to a little more detail. Without me having to.''}).

Participants also valued \revised{\textbf{time-lapse estimation}} with descriptions linked to actionable next steps. P7, after photographing sausages, imagined the system saying: \textit{``Maybe it still looks pretty pink—stir a little more, maybe about five minutes.''} This mirrors preferences in localization tasks, where relative distances (e.g., \textit{``a few steps to the left''}) helped guide actions. \revised{However, visual temporal understanding such as temporal order understanding and time-lapse estimation is limited in state-of-the-art VLMs~\cite{imam2025can}}. 

\subsubsection{\textcolor{black}{Identification tasks call for automatic verbosity control and personalization of proactive advice and camera guidance}}



Identification tasks centered on naming or recognizing objects, products, or items, often to decide how to use them or to distinguish between similar options. In these tasks, participants generally prioritized conciseness, favoring short, direct answers over lengthy descriptions. Unlike description tasks, identification in diary interactions often involved unfamiliar items, which shaped participants’ expectations of what counted as an ideal response.

In many cases, participants prioritized \textbf{binary confirmation} over lengthy output. Rather than full-scene descriptions, they wanted simple yes/no answers, such as whether a bag contained broccoli (P3), whether snow was on the grass (P5), or whether a park bench was empty (P5). Building on prior work~\cite{stangl2020person} that found the amount of time the user has influences the amount of content they want, some participants suggested a \textbf{“hurry mode”} toggle, providing only the bare minimum in time-sensitive contexts. As P11 explained about the interaction when he was getting ready to go out: \textit{``If you had a hurry button and I was in a hurry, that probably would be OK. If it just said no stain, it’s good to go.''}

At the same time, some participants welcomed \textbf{proactive advisory suggestions}. Clothing identification often invited proactive fashion advice (P10: \textit{``It could say, Well, that does match. But wearing a red shirt would look better.''}). Others, like P6, imagined instructions for use following identification, such as how to operate a product once recognized.

Another common need was \textbf{proactive image framing}. \revised{In most cases where participants struggled with image capture, the underlying issue was improper focus or cropping of the object. Several participants explicitly asked how to take a better picture, yet the system’s camera guidance remained generic—providing suggestions such as \textit{``ensure the room is well-lit,''} \textit{``hold the camera steady,''} or \textit{``adjust the angle to capture the full screen.''} which often did not address the primary problem. Participants expressed a clear need for more actionable, object-aware feedback. As P7 explained, \textit{``It would be useful if there was some kind of focus indicator… with rounded bottles or cans it tends to be harder to capture the image in a way that is interpretable for the AI… Just like when I’m taking a regular picture with the iPhone and it says ‘centered,’ ‘tilt left,’ ‘tilt right.’ Tools like that would help take better pictures to get better responses.''}}


Finally, several participants emphasized the need for \textbf{automatic verbosity control}, where the system could remember preferences and adapt detail level across contexts. As P7 noted, \textit{``In an ideal AI situation, where it does learn your preferences... it would still probably need some adjustment in different scenarios, but for the most part it could possibly get.''}

\subsubsection{\textcolor{black}{Text reading tasks call for personalization of content priorities and hierarchical summaries}}






These tasks involved reading information from product labels, menus, screenshots, and other written content. In the lab, participants consistently stated that the system provided too much information when reading entire ingredient lists on product packages. Almost all aparticipants expressed preference towards specifying question before, or along with the image input, allowing them to clarify what they wanted in advance. As P10 explained: \textit{``Even before I can take a picture, I could ask the question, how much protein is in this product? So that I don’t have to listen to all this other stuff that I don’t really care about.''} Prompting techniques like \textbf{intention prompting}, introduced in the lab, could similarly help streamline outputs.

In diary interactions, additional themes surfaced. Similar to description tasks, participants expressed preference for a \textbf{hierarchical response style}, starting with a high-level overview or summary and then offering the option to drill down for details. This mirrors screen reader conventions, where users skim at a fisheye view before zooming in. They also believed this approach could reduce errors from the system missing content. For example, P9, who tried to read a large restaurant menu, found that sections were repeatedly omitted. She suggested: \textit{``It could have listed the sections, appetizers, entrees, desserts, and then asked what I wanted to know about. That minimizes the risk of missing sections.''} She added that a high-level overview \textit{``would have been nice to know what all of them were, instead of having to ask several times.''} Related prompting techniques like \textbf{decomposition}, breaking the task into subtasks (e.g., reading menus section by section), could also support this goal.

Participants further emphasized the importance of \textbf{comprehensive product knowledge}, especially when reading text on packages. As P5 put it: \textit{``It would have been good to put everything there [in the database], so that when you have a product, maybe you take a picture, you can ask from that picture many information without taking several pictures.''}

\revised{\subsection{Participants Trust and Verification Practices}}
\subsubsection{Localization}
\textbf{Trust} in localization was generally lower than in other tasks. Participants were cautious about relying on the system for navigation and distance judgments, especially under time pressure. P2 did not trust it for \textit{``moving from one area to another,''} and while they were willing to use it alone in daytime or in unfamiliar areas, they avoided doing so at night. Safety concerns further limited trust; for instance, P1 preferred human help when cones blocked her path and P3 explained she would not use it in the street: \textit{``I wouldn't feel comfortable (to use it) on the street. because it would take away from my awareness of my surrounding''}

\textbf{Verification} also shaped localization practices. While some participants confirmed responses by physically moving toward and touching objects, some other described a broader trial-and-error process. For example, after using her cane to locate an exit door, P8 reflected: \textit{``It’s all trial and error…you have to use your cane and problem-solving skills to figure out, hey, did the app tell me correctly, did I do something wrong?''} She described this as a collaborative effort: \textit{``It’s a tag team effort, you know, me and the app are working on this together. It’s my brain and the AI.''} Rather than expecting perfect accuracy, participants saw the system as a partner in iterative problem solving.

\subsubsection{Description}
\textbf{Trust} in description tasks was generally tied to task and condition. Participants trusted static, factual details (P6: room layout; P5: when the picture was clear). P7 found it robust across different lighting conditions. Participants were more cautious with nuanced or subjective judgments. For instance P6 stated that for measurements she \textit{``couldn't rely on it''}. P11 trusted big-scene descriptions and explained: \textit{``I feel like I can trust it to tell me.
The big scene, like if a car in my driveway''}. P10 stated that color descriptions were not trusted at first try and sometimes requeried for consistency: \textit{``for colors, I would probably ask it a couple of times to make sure. I keep getting the same response. especially with, blacks, grays, and navies stuff like that.''}.

\textbf{Verification} in description tasks often involved \textbf{cross-checking with other AI tools}. For example, participants compared answers across apps when accuracy was important. One prompting approach that could support this is \textbf{ensembling}, where multiple prompts are used to solve the same problem and responses are aggregated into a final output. More broadly, participants’ practices highlight opportunities for systems to provide built-in verification, either by running additional checks in the background or by surfacing alternative answers, as emerging general AI tools now do.

\subsubsection{Identification}
\textbf{Trust and verification} in identification tasks was closely tied to stakes. For low-stakes scenarios, confidence was high and concise or binary answers were preferred. P3 trusted it for food and clothing, P7 for classification tasks like cars, P10 for road signs or sidewalk clearance, and P5 assumed the app would only report what was on the package, not invent details. For high-stakes items, especially medication, participants deferred to sighted assistance rather than risk error (P3, P7). Confidence increased when images were clear (P5) or when repeated queries produced consistent results (P10). Some novices displayed a \textbf{priori trust} in the system, as with P8, who was using this kind of VLM based VQA system for the first time. She explained: \textit{``I don't have a problem trusting AI. especially because of the results that I'm seeing. I think that AI just needs to remain very brutally honest, because that's what's going to make it maintain a reputable reputation.''} This echoes prior reports of high self‑reported trust and acceptance without verification in early interactions with multimodal LLM visual assistants, before users develop awareness of common error patterns~\cite{Meng2025surfacing, ricardo2025towards}.

\subsubsection{Reading text}
How participants expressed \textbf{trust} in these scenarios was mixed: P10 trusted OCR, P11 did not. P5 trusted package text under the assumption that \textit{``the app would not invent information.''} which can tie to \textbf{priori trust} mentioned in description tasks.
\textbf{Verification} in text reading tasks was less common compared to localization or description. Participants generally trusted textual outputs if they aligned with expectations. But in high-stakes cases, such as medication labels, test dates, or medical reports, they preferred sighted confirmation or cross-verifying across multiple AI tools rather than relying solely on one AI.

\section{Discussion}
This study provides an in-depth look at how blind users personalize and control a VLM-based conversational VQA system in both lab and real-world settings. By analyzing 418 interactions, plus participant reflections, and post-study interviews, we highlight how users actively shape system outputs, adapting prompts, verifying responses, and expressing preferences for verbosity and task specific guidance. In this section we situate our contributions within broader developments in the field.


\subsection{Implications}
\subsubsection{Inclusive Datasets}
Existing assistive VQA datasets have laid important foundations by grounding research in real images captured by blind and low-vision users. For example, VizWiz~\cite{vizwiz} provides data on short, single-turn question–answer interactions, and followup work has extended this line with visually grounding answers~\cite{chen2022grounding} and long-form responses that may contain information beyond the question answer such as explanations and suggestions~\cite{mina2024Long}. However, these datasets are still primarily limited to one-image, one-question interactions. Recent work has begun to explore multi-turn interactions relevant to assistive contexts, such as camera guidance when information in a single image is insufficient~\cite{liu2024right}, but they do so without releasing multi‑turn conversations grounded in images taken by blind people in real world context.

Our dataset extends this space by providing multi-turn conversational VQA collected with 11 blind participants using \textit{Be My AI}, spanning both controlled lab sessions and real-world use. It includes user-captured images, dialog transcripts, context-of-use metadata, and task success. 
This enables analysis of follow-ups, referential disambiguation, and perceptions of helpfulness in assistive scenarios. In doing so, it bridges assistive VQA datasets with the broader multi-turn visually grounded dialogue research, providing a blind-centered testbed for exploring emerging interactive behaviors.


\subsubsection{Relevance to Emerging Live VQA}
Our findings on image-based VQA personalization offer insights that can inform the development of the next generation of assistive video-based or live VQA systems. Recent efforts such as ChatGPT’s Advanced Voice with Video for real-time assistance to blind users~\cite{chang2025probing}, the EgoBlind dataset of egocentric video QA~\cite{xiao2025egoblind}, and prototype systems like VocalEyes~\cite{Chavan2024VocalEyes} and AIris~\cite{brilli2024airis} show growing interest in real-time assistance for blind users. However, their continuous frames and faster-paced interaction, heighten, rather than reduce, the need for accurate, user-tailored, and concise responses. The customization practices we document, such as toggling verbosity under time pressure, grounding spatial descriptions in the user’s perspective, offering proactive capture guidance, and supporting verification for high-stakes tasks, can inform and support the design of more effective live QA systems.


\subsubsection{Toward Teachable Generative AI}
Our findings indicate a strong interest among blind participants in persistent, user-driven personalization. Many wanted the system to remember preferred detail levels (e.g., fine-grained color shades or specific cooking cues) and to consistently provide task-oriented guidance across sessions. Existing tools with memory features, such as ChatGPT and Claude, can store and recall textual preferences, but our results suggest extending these capabilities beyond text to multimodal input. Building on earlier teachable machine concepts~\cite{findmything, kacorri2017people, Sosa2017hands}, users could label household items once, and have those preferences persist, supporting user-driven concept learning, automatic conditioning on verbosity or ordering of information, and adaptation to new environments while retaining individual guidance. While prototypes of memory-enabled LLMs exist, no widely deployed system yet offers blind users a seamless multimodal way to \textit{``teach''} generative AI, pointing to a direction for future assistive technology.

\subsection{Limitations}

\revised{Several factors related to the platform used for data collection shape the limitations of this work. All interactions were conducted using Be My AI at a fixed stage of its development. We selected this application because, at the time of the study, it was the only screen-reader–accessible, real-world VQA tool suitable for deployment with blind participants. While Be My AI was not designed for research or for persistent customization, its accessibility and widespread use made it an appropriate platform for observing authentic interaction patterns. Because the system did not retain user preferences across sessions, participants’ adaptations could only influence the current dialog.}

\revised{To help surface the kinds of stable preferences participants wished an ideal system would remember, we asked them to use customization techniques as if persistence were available. This approach aligns with prior work showing that effective personalization depends on a system’s ability to store and reuse user-relevant information over time \cite{zhong2024memorybank, park2023generative}. The lack of such persistence in our deployment context likely shaped how participants experimented with customization during the diary study. For instance, P9 frequently asked about color, marbling, and marinade distribution when photographing meat—preferences she hoped a future system could remember: \textit{“If I said the first time that I want marbling and color, then ideally it should remember that for the next time.”} Such cases illustrate how participants adapted to the system’s current capabilities while articulating expectations for future, more persistent forms of customization.}

\revised{These platform characteristics also informed our study duration. Extending the deployment substantially longer would likely have increased fatigue, as participants would need to repeat the same preferences without any system retention. A ten-day period was deemed long enough for recurring scenarios and customization opportunities to emerge while limiting this potential burden.}

\section{Conclusion}
In this study, we investigate how blind users exert control over AI-generated responses in conversational VQA systems through a range of prompting-based customization techniques. Through a three-phase study combining guided in-lab session and in-the-wild diary session, we find that short participant inputs elicit responses over ten times longer and can extend up to 21 turns, especially when users recapture images, clarify ambiguous outputs, or follow curiosity-driven questions.  
Participants actively shaped responses using both introduced and self-invented customization techniques, to manage verbosity, focus on task-relevant details, and strengthen trust. These strategies appeared across professional, familiar, and unfamiliar contexts, with slightly higher rates in familiar settings.  
Reflections highlight key design needs for future assistive VQA systems: on-demand verbosity control, user-centered spatial references, proactive camera guidance, and verification mechanisms for high-stakes tasks.  
We publicly release \datasetname, our dataset capturing blind users’ real-world and in-lab interactions with Be My AI, along with their reflections and detailed context of use. Together, these contributions point toward a future of assistive AI that strengthens blind users’ autonomy by enabling richer customization, supporting individual goals, preferences, and everyday contexts.

\begin{acks}
We thank the blind participants in our study and The National Federation of the Blind for their assistance with participant recruitment. The contents of this paper were developed under grant from the National Institute on Disability, Independent Living, and Rehabilitation Research (NIDILRR Grant No. 90REGE0024). NIDILRR is a Center within the Administration for Community Living (ACL), Department of Health and Human Services (HHS). The contents of this paper do not necessarily represent the policy of NIDILRR, ACL, HHS, and you should not assume endorsement by the Federal Government. This material is based upon work partially supported by the NSF under Grant No. 2229885 (NSF Institute for Trustworthy AI in Law and Society, TRAILS). Any opinions, findings, and conclusions or recommendations expressed in this material are those of the author(s) and do not necessarily reflect the views of the National Science Foundation.
\end{acks}

\bibliographystyle{ACM-Reference-Format}
\bibliography{sample-base}

\appendix
\section{Introducing Customization Techniques in the Lab}\label{appendix:A}
\textbf{Scenario 1: Computer Screen Error Message - No Technique.}
The first scenario task is using the system to interpret an error message on a laptop screen. This read-text task appears frequently in the VizWiz dataset and was validated by the pilot researcher as a task he would personally use \textit{Be My AI} for. This initial task serves to familiarize participants with the system’s core functionality while giving them a chance to practice sharing their interaction session with the research team. To avoid cognitive overload, we do not introduce any customization techniques during this task.

\textbf{Scenario 2: Exit Door Navigation - Binary Feedback and Style Prompting.}
In the second scenario, participants are asked to navigate to an exit door in an unfamiliar hallway using the system. Many studies have investigated approaches to assist blind individuals with navigation tasks \citep[e.g.][]{Islam_2024, Tang_2025}. After completing the interaction, participants are introduced to the binary feedback technique, which they can use to provide simple responses--such as \textit{``helpful''} or \textit{``not helpful''}--to indicate their preferences. This technique is framed as a way to signal to the system what kinds of responses they found useful, imagining that future responses might be shaped accordingly.
We also introduce participants to style prompting, which enables them to explicitly guide the format or tone of the system’s responses. For example, they might instruct the system to \textit{``be less verbose''} or \textit{``give more details when giving directions.''}

\textbf{Scenario 3: Nutrition Facts - Intention Prompting.}
In this task, participants are asked to find the nutrition information on a Lipton soup mix box. Before beginning, they are encouraged to set a specific goal for the task, such as monitoring sugar intake and determining whether the product is suitable for their diet. Tasks involving nutritional content also frequently appear in the VizWiz dataset. The technique introduced in this scenario is intention prompting, in which participants include their goal as part of their query such as \textit{``What is the nutrition of this pacakge? I'm trying to control sugar intake.''} An ideal system would use this context to tailor its response—omitting irrelevant details (e.g., calcium content) and focusing on information aligned with the user’s stated intention. This approach also reduces the need for users to ask separate questions, as the system could provide all relevant details based on the intention.

\textbf{Scenario 4: Item Localization - Chain of Thought and Decomposition Prompting.}
In the fourth scenario, participants are asked to find the same box of soup used in the previous task, now placed somewhere within the lab. This task focuses on localization and navigation. 
Rather than using the typical Chain-of-Thought (CoT) prompting phrase: \textit{``Let’s think step by step,''} we introduce a more natural way for participants to prompt with structured reasoning. For example, a participant can say, \textit{``I only care about locating the soup, not the other items in the scene like the painting on the wall.''} With explicit reasoning and instruction, an ideal system would be better in generating responses that align with users' informational needs. In diaries, we also observe participants naturally using a series of decomposed questions to work toward a solution (see \autoref{sec:4.2}). This pattern resembles both CoT prompting and decomposition prompting. Given the overlap in purpose and function \citep{schulhoff2025promptreportsystematicsurvey}, we group these prompting techniques together in our analysis.

\textbf{Scenario 5: Personal Item Localization - SimToM Prompting.}
The final scenario involves locating a personal item of the participant’s choice (e.g., keys) that has been deliberately misplaced in the lab room. Locating personal items is a common real-world use case for blind users of visual assistance systems \citep{findmything}. In this task, participants are introduced to SimToM (Simulate the Thought of a Mind) prompting, a technique that encourages them to guide the system by describing facts they would use to approach the task themselves. Here, participants are instructed to describe the item’s features (e.g., shape, color, or attached objects) to help the system better understand what to look for. By incorporating this kind of explicit reasoning and instruction, the system is expected to effective and accurate responses. 

\section{Diary Study Practice Days Protocol}\label{appendix:B}
On Day 1, participants are asked to try using the system to identify everyday objects within their home, such as food or clothing. On Day 2, they are encouraged to explore how lighting conditions may affect the system’s performance by taking images in natural daylight, with poor lighting conditions, and with artificial lighting. They are also encouraged to try the system with accessibility-related objects (e.g., canes). On Day 3, participants are asked to step outside of familiar spaces and try the system in new or social settings, such as outdoors or around unfamiliar people. Participants are instructed to complete at least three interactions per day during the three-day practice session.

\section{End-of-Day Diary Study Reflection Questions}\label{appendix:C}
During the diary study, participants received a set of reflection questions at the end of each day, including: (1) Was the system helpful today? (2) Did you try to give it any feedback or make it work better for you? (3) How could you tell if the system was right or wrong? and (4) Is there anything else you would like to add? Responses to these questions, alongside the interaction logs, serve as key data sources in our analysis of real-world usage and customization techniques.

\section{Our categorization on the types of questions asked by participants}
See table \ref{tab:question_types}
\begin{table*}[b]
\caption{Types of questions asked by participants, with definitions and examples.}
\small
\centering
\renewcommand{\arraystretch}{0.92}
\begin{tabular}{p{12em} p{14em} p{18em}}
\toprule
\textbf{Question Types} & \textbf{Definition} & \textbf{Example Questions} \\
\midrule

\mbox{Description Questions} \newline (40.38\%\textsuperscript{*}) &
Questions asked for the description of a picture or the visual characteristics of an object or a scene. &
``Is anyone sitting on the bench?'' \newline
``What color are the pillows on the couch?'' \newline
``Please just describe in the greatest possible detail the picture [on the screenshot].'' \\
\addlinespace

\mbox{Reading Questions} \newline (20.33\%) &
Questions asked for reading text or looking for information based on text. &
``I need to know who the sender is.'' \newline
``What is written on the door?'' \newline
``May you read me the setup instructions?'' \\
\addlinespace

\mbox{Localization Questions} \newline (9.62\%) &
Questions asked for the location or spatial relation of objects. &
``Where is the ID card?'' \newline
``Help me find the remote.'' \newline
``From where I am, how far is the dumpster and is it to my left or right?'' \\
\addlinespace

\mbox{Identification Questions} \newline (8.79\%) &
Questions asked for identifying an object by name or type\cite{Brady2013investigating}. &
``What is it?'' \newline
``What is in the boxes?'' \newline
``What's the Stonyfield's product? Is it yogurt'' \\
\addlinespace

\mbox{Unanswerable Questions} \newline (5.49\% unanswerable till the end) &
Questions that cannot be answered based on the information included in previous images in the interaction. &
-- \\
\addlinespace

\bottomrule
\end{tabular}
\begin{minipage}{0.87\textwidth}
    \smallskip
    \textsuperscript{*}Percentage is calculated from all diary study interactions that include questions of this type.
\end{minipage}

\label{tab:question_types}
\end{table*}

\end{document}